\documentclass{aa}  
\bibpunct{(}{)}{;}{a}{}{,} 
\usepackage{graphicx}
\usepackage{txfonts}
\usepackage{xcolor}
\usepackage{epsfig}
\usepackage{amsmath}
\usepackage{natbib}
\usepackage{bm}
\usepackage{physics}
\usepackage{subfigure}
\usepackage[normalem]{ulem}
\usepackage{hyperref}
\newcommand{\rev}[1]{\textcolor{red}{#1}}

\newcommand{\apo}[1]{}
\newcommand{\me}[1]{\rev{\sout{We} I}}
\newcommand{\lme}[1]{\rev{\sout{we} I}}

\begin{document}

   \title{DISPATCH methods: an approximate, entropy-based Riemann solver for ideal magnetohydrodynamics}
   
   \authorrunning{Popovas}
   \titlerunning{An approximate, entropy-based Riemann HLLD solver for ideal MHD}

   \author{Popovas Andrius
 		 \inst{1,2}
 		 }

   \institute{Rosseland Centre for Solar Physics, University of Oslo, P.O. Box 1029, Blindern, NO-0315 Oslo, Norway\\
\email{andrius.popovas@astro.uio.no}
\and
Institute of Theoretical Astrophysics, University of Oslo, P.O. Box 1029, Blindern, NO-0315 Oslo, Norway
}

\date{Received; accepted}
  \abstract{
   {With the advance of supercomputers we can now afford simulations with very large ranges of scales. In astrophysical applications, e.g. simulating Solar, stellar and planetary atmospheres, interstellar medium, etc; physical quantities, like gas pressure, density, temperature, plasma $\beta$, Mach, Reynolds numbers can vary by orders of magnitude. This requires a robust solver, which can deal with a very wide range of conditions and be able to maintain hydrostatic equilibrium where it is applicable.}

   {We reformulate a Godunov-type HLLD Riemann solver that it would be suitable to maintain hydrostatic equilibrium in atmospheric applications in a range of Mach numbers, regimes where kinetic and magnetic energies dominate over thermal energy without any ad-hoc corrections.}

   {We change the solver to use entropy instead of total energy as the primary thermodynamic variable in the system of MHD equations. The entropy is not conserved, it increases when kinetic and magnetic energy is converted to heat, as it should.}

   {We propose using an approximate entropy - based Riemann solver as an alternative to already widely used Riemann solver formulations where it might be beneficial. We conduct a series of standard tests with varying conditions and show that the new formulation for the Godunov type Riemann solver works and is promising.}}

   {}


   \maketitle
   
\section{Introduction}

Turbulent magnetised gas is ubiquitous in astrophysics. In a single system (e.g. stellar and planetary atmospheres, interstellar medium, etc.), the physical quantities, like density, gas pressure, entropy and temperature can vary by orders of magnitude. Let's take our closest star, the Sun, as an example. It has several distinct regions. In the core, energy is generated by nuclear fusion and diffused outward by radiation (gamma and X-rays). When the temperature drops sufficiently enough for ions to recombine the opacity of the plasma increases and the radiative diffusion can no longer efficiently transport the energy outwards, which causes the plasma to start to simmer. Here, the convection zone begins. The convection zone spans roughly 200 Mm and makes up the outer 30\% of the solar interior. The mean free path of the photons increases as the density of the plasma decreases with increasing radius, eventually allowing the photons to escape. This is the optical surface that we can observe, the photosphere. Above the photosphere, in the chromosphere, the thermal energy of the gas no longer dominates the total energy budget, plasma $\beta$ drops below unity, indicating that magnetic pressure becomes higher than the gas pressure, and the gas flows must follow the magnetic field configuration. Even higher up, the gas temperature suddenly goes back up to millions degrees Kelvin. The physical quantities, like density, gas pressure, temperature, vary by many orders of magnitude from the deep interior all the way up to the outermost parts of the Sun. However, in the convection zone, radially averaged entropy per unit mass is nearly constant. This suggests using entropy as one of the main quantities in solar and stellar simulations. Numerically maintaining hydrostatic equilibrium is much easier when a quantity varies very little with depth, unlike temperature or internal/total energy. There are already works done, where entropy is used as a primary thermodynamic variable, e.g. works by \cite{Popovas_2018,Popovas_2019,Hotta_2022}, however the partial differential equation (PDE) solvers in these works are either very diffuse in low velocity motions (i.e. \citealt{Popovas_2018,Popovas_2019}) or treat entropy as a conservative quantity and don’t account for entropy increase in shocks.

With the advent of exascale computing we are able to run numerical simulations of increasing complexity - either covering larger range of physical and temporal scales, or including more physics, like radiative heat transfer, conductivity, chemistry, non-local thermal equilibrium, generalized Ohm's law, etc. A very popular group of tools to solve the equations of magnetohydrodynamics (MHD) is a Godunov-type Riemann solvers, e.g.  HLL, HLLC, HLLD, Roe solvers \citep{Roe1981,Harten_1983,Toro_1994,Miyoshi_2005,Teyssier_2002,Teyssier_2006,Gurski_2004,Li_2005,Toro_2019,Winters_2016,Derigs_2016} and references within. These solvers are efficient in conditions where detailed turbulent structures ought to be preserved, but due to numerical precision cannot accurately evolve thermal energy, when kinetic and magnetic energies dominate the energy budget -- this can create negative pressure spots in  e.g. Solar corona, cause inaccurate convective flow velocities, or create areas with negative temperature in a giant molecular cloud simulations. This is a well known problem with HLL and Roe type Riemann solvers (e.g. \citealt[section 6.6]{Teyssier_2015}). Many attempts have been made to mitigate the issue (e.g. \citealt{Ryu_1993, Bryan_1995,ismail_2009,Winters_2016,herbin_2020,Gallice_2022}), which do not necessarily resolve the problems, but at least make them negligible under certain conditions. One alternative to these fixes would be using logarithmic density and pressure.

In this work we present a prescription for an approximate entropy-based Riemann solver, which for simplicity here we will call HLLS. It is based on the HLLD solver in RAMSES code \citep{Fromang_2006}, also see \cite{Miyoshi_2005} for a detailed description of a HLLD solver. The HLLS solver is \textbf{not entropy conserving} -- shocks and magneto-acoustic waves generate entropy by converting kinetic and magnetic energy to heat. It is an irreversible process -- in a closed system energy cannot be created or destroyed; it can only change its form. This is a fundamental requirement and is mathematically\footnote{but not necessarily numerically.} fulfilled in total energy-based Riemann solvers. As we are converting the solver to an entropy-based one, we must take into account the entropy increase in shocks and magneto-acoustic waves. In principle, the generated entropy $S_{gen}$ depends on the state of the Riemann fan and is characterised by the change in primitive quantities - density, velocity and total pressure, when crossing a given discontinuity. But in practice, computing entropy production via discontinuities is insufficient. Riemann solvers are diffusive due to the use of slope limiters as well as finite numerical precision.
Section \ref{sec:tot_e_riemann_solver} will briefly go through the key aspects of the HLLD Riemann solver, as later they will be relevant when discussing the changes and underlying considerations when converting to an entropy-based solver in section \ref{sec:hlls_solver}. We briefly discuss the equation of state where entropy is one of the primary variables and how to compute one in section \ref{sec:S_EOS}. We continue by showing a few numerical tests in section \ref{sec:numerical_tests}. In section \ref{sec:discussion} we discuss the current limitations of the solver and the applications, where this solver would be the most beneficial. We conclude by discussing our future development work.

\section{The total energy-based HLLD Riemann solver} \label{sec:tot_e_riemann_solver}

The ideal MHD equations are a classical system of equations used to study the dynamics of conducting fluids. In traditional Riemann solvers the fluid equations for the total mass density, momentum density, total energy density and the magnetic field of the system can be written as a system of conservation laws,

\begin{equation}
    \frac{\partial}{\partial t} 
    \left[ \begin{array} {c} \rho \\ 
    \rho \boldsymbol{u} \\ 
    E_{\rm{tot}} \\ 
    \boldsymbol{B} \end{array} \right] +
    \nabla \cdot \left[ \begin{array} {c} \rho \boldsymbol{u} \\ 
    \rho (\boldsymbol{u} \otimes \boldsymbol{u}) + P_{\rm{tot}} - \boldsymbol{B} \otimes \boldsymbol{B} \\
    \ \boldsymbol{u} (E_{\rm{tot}} + P_{\rm{tot}}) - \boldsymbol{B}(\boldsymbol{u} \cdot \boldsymbol{B}) \\
    \boldsymbol{B} \otimes  \boldsymbol{u} - \boldsymbol{u} \otimes \boldsymbol{B} \end{array} \right] = 0,
\label{eq:basic_eq_set_tot_E}
\end{equation}
with
\begin{equation}
    \nabla \cdot \boldsymbol{B} = 0,
\end{equation}

where $\rho$, $\boldsymbol{u}$, $E_{tot}$, $P_{tot}$, and $\boldsymbol{B}$ are density, velocity, total energy, total pressure, and magnetic field respectively. Total energy is a sum of internal, kinetic and magnetic energies,
\begin{equation}
    E_{\rm{tot}} = \rho \varepsilon + \frac{\rho \boldsymbol{u}^2}{2} + \frac{|\boldsymbol{B}|^2}{2},
\end{equation}
where $\varepsilon$ is internal energy per unit mass; and the total pressure $P_{\rm{tot}}$ is a sum of gas and magnetic pressures,
\begin{equation}
     P_{\rm{tot}} = P_{\rm{gas}} + \frac{|\boldsymbol{B}|^2}{2}.
\end{equation}
The system is closed by an equation of state (EOS). In this work we use a simple gamma-law (ideal gas) EOS:
\begin{equation}
\label{eq:eos_eth}
    P=\rho (\gamma - 1) \varepsilon,
\end{equation}
where $\gamma$ is the adiabatic index. 

The beauty of the system \ref{eq:basic_eq_set_tot_E} lies in the total energy - when one type of energy is converted to another, one does not have to explicitly compute the conversion via e.g. Maxwell stress tensors, Lorentz force, diagonal Reynolds stress, etc. It saves time and retains very high precision, at least in theory. In practice, however, there are complications, mostly due to numerical floating point precision. If magnetic or kinetic energy strongly dominates over thermal energy, the precision of the latter drops to an extent that the expressed pressure or temperature can become negative.

\section{The entropy - based equation of state}
\label{sec:S_EOS}
To define the temperature and entropy,  we use ideal gas law,
\begin{equation}
    \label{eq:ideal_gas_law}
    P = \frac{\rho k_B T}{\mu m_u},
\end{equation}
where $k_B$ is the Boltzmann constant, $\mu$ is the mean molecular weight, $T$ is the gas temperature, and $m_u$ is the atomic mass unit. Entropy per unit mass is defined as
\begin{equation}
\label{eq:basic_S}
    S = S_0 + c_v ~\ln(P\rho^{-\gamma}),
\end{equation}
where $c_v= \left(\frac{\partial \varepsilon}{\partial T}\right)_\rho$ is the heat capacity at constant volume. Here we introduce the entropy null-point $S_0$. It can be considered as an integration constant or the starting point of the summation of the microstates. Having $S_0$ is not necessary (i.e. $S_0=0$) and mathematically entropy value can be both positive and negative, but physically entropy should not go below 0\footnote{Entropy is usually considered to be either always positive or always negative, or normalised within an interval of $[0,1]$, with 0 marking the lowest level of disorder.}. From a computational point of view it is best to keep $S$ as close to zero as possible -- then small gradients in entropy retain the precision.

It is really convenient to express eq. \ref{eq:basic_S} as
\begin{equation}
\label{eq:basic_S2}
    S =  S_0 + \frac{k_B}{(\gamma - 1) \mu m_u} ~\ln(P\rho^{-\gamma}) = S_0 + \frac{\wp}{\gamma-1} ~\ln(P\rho^{-\gamma}),
\end{equation}
where $\wp = \frac{k_B}{\mu m_u} = R$, where $R$ is the gas constant. Now we can notice that $\wp$ repeats itself in both the entropy definition and the ideal gas law. Thus, we can use it as a scaling factor for both temperature and entropy, making the modified entropy,
\begin{equation}
\mathfrak{S}(\gamma-1) = S_0 + ~ln(P\rho^{-\gamma}),    
\end{equation}
a dimensionless quantity, which is also much closer to unity, thus making numerical operations with it more accurate. Converting to code-units is a common practice in numerical simulations.

Entropy in eq. \ref{eq:basic_S} represents the thermodynamic concept of entropy, which is a measure of the disorder or randomness of a system. For more realistic gases entropy per mole can be expressed from total partition sums (e.g. \citealt[eq. 43]{Popovas_2016}):
\begin{equation}
S = R~ \ln{Q_{\rm{tot}}} + \frac{RT}{Q_{\rm{tot}}}\left( \frac{\partial Q_{\rm{tot}}}{\partial T} \right),
\end{equation}
where $Q_{\rm{tot}}$ is the total partition sum.

For arbitrary tabular EOS, entropy can be derived from other quantities by integrating
\begin{equation}
\label{eq:thermodynamic_S}
  dS = \left[ \frac{1}{T} \left( \frac{\partial \varepsilon}{\partial T}  \right)_\rho \right] dT + \left[ \frac{1}{T} \left( \frac{\partial \varepsilon}{\partial \ln\rho}  \right)_T - \frac{P}{T \rho}\right] d\ln\rho.
\end{equation}
The first term in the second brackets on the RHS of the eq. \ref{eq:thermodynamic_S} is generally very small and for ideal gas vanishes completely, leading to
\begin{equation}
\label{eq:simplified_thermodynamic_S}
  dS = \left[ \frac{1}{T} \left( \frac{\partial \varepsilon}{\partial T}  \right)_\rho \right] dT - \left[ \frac{P}{T \rho}\right] d\ln\rho.
\end{equation}
It is however very important to check that the new entropy-based table satisfies the Maxwell's thermodynamic identity,
\begin{equation}
\label{eq:maxwell_identity}
    \left( \frac{\partial \varepsilon}{\partial \rho} \right)_T = \frac{P}{\rho^2} - \frac{T}{\rho^2} \left( \frac{\partial P}{\partial T} \right)_\rho = \frac{P}{\rho^2} \left[ \left( \frac{\partial \ln{P}}{\partial \ln{T}}  \right)_\rho  \right],
\end{equation}
and gives exact results to numerical precision for requested quantities, when compared to the original table. In fact, any EOS must satisfy the thermodynamic identity \ref{eq:maxwell_identity}.

Lastly, sometimes entropy is adopted to be $S = P\rho^{-\gamma}$, e.g.\ \citealt{Ryu_1993,Cuissa_2022}, called  modified entropy, which should be considered as a different kind of polytropic EOS and doesn't have a real physical meaning. Although this quantity could be used as a probe for thermal energy, it does not relate to thermodynamic entropy. On the other hand, it works really well maintaining thermal energy stability in ideal gases, as shown by \cite{Ryu_1993}. It is not surprising, since the quantity is advected as a passive scalar, it will be conserved to numerical precision despite not giving accurate values where entropy (or thermal energy in this case) should increase.

\section{The conversion to an entropy-based HLLD solver}
\label{sec:hlls_solver}

In DISPATCH \citep{Nordlund_2018MNRAS} implementation of this new solver we use the MUSCL-Hancock algorithm with constrained transport \citep[CT]{Evans_1988} for the induction equation, as well as a positivity preserving 3D unsplit TVD slope limiter; see \citet{Fromang_2006} for details. For convenience, we will not go through the full details of the Godunov scheme and will use standard notation. However, we summarise the modified numerical method in appendix \ref{app1:numerical_method} with the changes depicting the conversion from total energy to entropy. For a deeper overview about HLL, HLLC and HLLD Riemann solvers we refer to \cite{Miyoshi_2005,Toro_2019} and references therein. The main difference from typical HLLD solvers are just the parts where total energy is replaced with entropy with additional fluxes computed to calculate the entropy production (details below).

In the system of equations \ref{eq:basic_eq_set_tot_E} we replace the total energy term with entropy per unit mass:

\begin{equation}
    \frac{\partial}{\partial t} \left[ \begin{array} {c} \rho \\ \rho \boldsymbol{u} \\ \rho S \\ \boldsymbol{B} \end{array} \right] +
    \nabla \cdot \left[ \begin{array} {c} \rho \boldsymbol{u} \\ \rho (\boldsymbol{u} \otimes \boldsymbol{u}) + P_{\rm{tot}} - \boldsymbol{B} \otimes \boldsymbol{B}   \\ \rho \boldsymbol{u} S \\ \boldsymbol{B} \otimes  \boldsymbol{u} - \boldsymbol{u} \otimes \boldsymbol{B} \end{array} \right] = \left[ \begin{array} {c} 0 \\ \boldsymbol{\Phi}\\ (\boldsymbol{Q_{\rm{ext}}} + \boldsymbol{Q_{S}})/T \\ 0 \end{array} \right].
\label{eq:basic_eq_set_S}
\end{equation}

For completeness, here we also show the additional source terms, e.g. $\Phi$, force per unit volume, which includes force of gravity, Coriolis force, etc; $\boldsymbol{Q_{ext}}$ is external heating per unit volume, which includes Newton cooling, radiative heat transfer, etc; $\boldsymbol{Q_{S}}$ is entropy production from converting kinetic and magnetic energy to heat. The source terms from gravity, Coriolis forces, Newton cooling, etc. are added during both the predictor and the corrector steps.

The system in \ref{eq:basic_eq_set_S} can be written in vectorial form, similar to e.g. \cite{Londrillo_2000},
\begin{equation}
    \frac{\partial \bf{U}}{\partial t} + \frac{\partial \bf{F}}{\partial x} + \frac{\partial \bf{G}}{\partial y} + \frac{\partial \bf{H}}{\partial z} = \Psi,
\end{equation}
where
\begin{equation}
    \bf{U} = (\rho, \rho \rm{u}_x, \rho u_y, \rho u_z, \rho S, B_x, B_y, B_c)^T,
\end{equation}
\begin{equation}
    \Psi = (0, \Phi_x, \Phi_y, \Phi_z, Q/T + Q_{S}/T, 0, 0, 0)^T,
\end{equation}
and
\begin{equation}
\label{eq:basic_flux}
    \bf{F} = \left( \begin{array}{c} \rho u_x \\
    \rho u_x^2 + P_{tot} - B_x^2 \\
    \rho u_x u_y - B_x B_y \\
    \rho u_x u_z - B_x B_z \\
    \rho u_x S \\
    0 \\
    u_x B_y - u_y B_x \\
    u_x B_z - u_z B_x
    \end{array}
    \right)
\end{equation}
is the flux function. The expressions for the terms $\bf{G}$ and $\bf{H}$ are completely analogous. These equations have seven eigenvalues, corresponding to four magneto-acoustic (two slow and two fast), two Alfv\'en waves and an entropy wave:
\begin{equation}
\label{eq:7_waves}
    \lambda_{1,7} = u \mp c_f,  \\ \lambda_{3,5} = u \mp c_s, \\ \lambda_{2,6}=u \mp c_a,  \\ \lambda_4 = u,
\end{equation}
where
\begin{equation}
    c_{f,s}^2 = d  \pm \sqrt{d^2 - aB_x^2/\rho} , \\ c_a = \frac{|B_x|}{\sqrt{\rho}}
\end{equation}
with
\begin{equation}
    d = \frac{a^2 + |\boldsymbol{B}|^2/\rho}{2},
\end{equation}
and $a$ is the speed of sound. More often than not some eigenvalues in \ref{eq:7_waves} coincide,
\begin{equation}
    \lambda_1 \leq \lambda_2 \leq \lambda_3 \leq \lambda_4 \leq \lambda_5 \leq \lambda_6,
\end{equation}
depending on the direction and strength of the magnetic field. A solution to the Riemann problem then may be composed not only of ordinary shock and rarefaction waves, but also compound waves \citep{Brio1988400,Miyoshi_2005}. \\
In the set of equations \ref{eq:basic_flux} we do not show the additional fluxes that will be needed to compute the entropy generation, which is discussed in detail below.

\subsection{Entropy production}
\label{subsec:s_gen}
The entropy production in shocks from dissipation of kinetic energy has long been a topic of discussion \cite{Salas_1995}. Kelvin and Rayleigh questioned the validity of shock discontinuity as it violated the conservation of entropy. A very curious result was put forward by \cite{morduchow_1949}, that equilibrium entropy has a maximum inside a Navier-Stokes shock profile, indicating that entropy was decreasing after passing the shock, strengthening the doubts by Stokes, Kelvin and Rayleigh. \cite{Salas_1995} explored this curiosity and concluded that because of this phenomenon, the entropy propagation equation cannot be used as a conservation law. Using a single jump condition for propagation of entropy is not adequate, as going from equations with two jumps $[P]$ and $[u]$ to an equation with a single jump $[S]$ information is lost. See section 5 in \cite{Salas_1995} for details. Shocks in the infinitesimally small area cannot be considered equilibrium, as locally the adiabatic index changes because of the rapid compression. Rankine-Hugoniot jump conditions are only applicable when elements are far enough from the discontinuity to assume the local equilibrium is present on each of the both sides, and it is not resolved, what happens in the discontinuity. \cite{Margolin_2017} studies non-equilibrium entropy in a shock and shows, that it increases monotonically inside the shock and a certain modification can be done to the equilibrium formulation that it would follow the non-equilibrium formulation better. \cite{Thorbner_2008} derived analytical formulae for the rate of increase of entropy at arbitrary jumps in primitive variables for Godunov methods. It is then later used for total energy corrections.

Instead of correcting the total energy, we rewrite the Riemann solver to work in terms of entropy as a primary thermodynamic variable. We can split the evolution of entropy into two parts, the advection of the conserved entropy and new entropy production via dissipation of kinetic and magnetic energy. The conserved entropy is advected through the Riemann fan as a passive scalar, see appendix \ref{app1:numerical_method} for details. For the second part, we start by looking at the evolution of total, kinetic, magnetic and thermal energies. The total energy can be split into its constituent thermal ($E_{th}$), kinetic ($E_{kin}$) and magnetic ($E_{mag}$) energies,
\begin{equation}
    E_{\rm{tot}} = E_{\rm{th}} + E_{\rm{kin}} + E_{\rm{mag}},
\end{equation}
while it's evolution is then
\begin{equation}
    \frac{dE_{\rm{tot}}}{dt} = \frac{dE_{\rm{th}}}{dt} + \frac{dE_{\rm{kin}}}{dt} + \frac{dE_{\rm{mag}}}{dt}.
\end{equation}

Since the total energy is conserved, any local change in one form of energy must be accounted for by corresponding fluxes or conversions into other forms. To properly track entropy production, we examine the contributions from kinetic, magnetic, and thermal energies separately,
\begin{align} 
\label{eq:dekin_dt}
\frac{dE_{\rm{kin}}}{dt} &= -\div{\vb{F}_{\rm{kin}}} + W_{\rm{gas}} - Q_{\rm{kin}} + \Theta_{\rm{kin}}, \\
\frac{dE_{\rm{th}}}{dt} &= -\div{\vb{F}_{\rm{th}}} - W_{\rm{gas}} + Q_{\rm{kin}} + \Theta_{\rm{th}}, \\
\label{eq:demag_dt}
\frac{dE_{\rm{mag}}}{dt} &= -\div{\vb{F}_{\rm{mag}}} - \Theta_{\rm{kin}} - \Theta_{\rm{th}},
\end{align}
where $\div{\vb{F}_{\rm{kin}}}$,  $\div{\vb{F}_{\rm{th}}}$, $\div{\vb{F}_{\rm{mag}}}$ are the kinetic, thermal and magnetic energy fluxes respectively, $\frac{dE_{\rm{kin}}}{dt}$, $\frac{dE_{\rm{th}}}{dt}$ and $\frac{dE_{\rm{mag}}}{dt}$ are the change of kinetic, thermal and magnetic energies in a given cell in a time-step $dt$. $Q_{\rm{kin}}$ is heating from kinetic energy dissipation, $\Theta_{\rm{kin}}$ and $\Theta_{\rm{th}}$ are conversion of magnetic energy to kinetic and thermal energies respectively (e.g. reconnection, Ohmic dissipation, etc.). Lastly, $W_{\rm{gas}} = P \left(\div{\vb{u}}\right)$ is the pressure work.
From equations \ref{eq:dekin_dt} to \ref{eq:demag_dt} we can derive the heating from irreversible energy dissipation $\boldsymbol{Q_{\rm{S}}} = Q_{\rm{kin}} + \Theta_{\rm{th}}$:
\begin{equation}
\label{eq:q_s}
    \boldsymbol{Q_{\rm{S}}} =  -\div{\vb{F}_{\rm{kin}}} - \div{\vb{F}_{\rm{mag}}} - \frac{dE_{\rm{kin}}}{dt} - \frac{dE_{\rm{mag}}}{dt} + W_{\rm{gas}}.
\end{equation}
Therefore, to compute $\boldsymbol{Q_{\rm{S}}}$ we need additional quantities from the corrector step - the kinetic and magnetic energy fluxes as well as the normal velocities at the cell faces to compute $\div{\vb{u}}$. These velocities can be extracted from the corrector step by passing the normal velocity along with the fluxes. See \ref{app1:numerical_method} for the complete description of the corrector step. To accurately compute $\boldsymbol{Q_{\rm{S}}}$, a consistent evaluation of the kinetic and magnetic energy time derivatives is needed.

The time derivatives of kinetic and magnetic energy, $\frac{dE_{\rm{kin}}}{dt}$ and $\frac{dE_{\rm{mag}}}{dt}$, are computed by evaluating these quantities at both the old (at $t$) and new (at $t+1$) time steps. Specifically, $E_{\rm{kin}}$ and $E_{\rm{mag}}$ are first computed from the conserved variables at the beginning of the time step. Then, after updating the density, momentum, and magnetic fields via flux divergences, the new values of $E_{\rm{kin}}$ and $E_{\rm{mag}}$ are computed. The time derivatives are then approximated as simple finite differences between these two states. This approach ensures consistency with the numerical fluxes and avoids introducing additional diffusive or dispersive errors. Furthermore, this discretization is fully consistent with the MUSCL-Hancock predictor-corrector scheme used in the solver. If a different time integration scheme, such as a higher-order Runge-Kutta method, were employed, the computation of $\frac{dE_{\rm{kin}}}{dt}$ and $\frac{dE_{\rm{mag}}}{dt}$ would need to be adjusted accordingly to maintain consistency with the chosen temporal discretization.

This entropy generation calculation recipe fully accounts for entropy production in shocks and shearing, as well as for numerical dissipation of kinetic and magnetic energy via e.g. slope limiters, thus fully closing the system. Lastly, to fully adhere to the second law of thermodynamics, we require $\boldsymbol{Q_{\rm{S}}}$ to be positive definite:
\begin{equation}
\label{eq:S_update}
    \frac{\partial \left(\rho S\right)}{\partial t} = - \div{\left(\rho \vb{u} S\right)} + max\left(0,\frac{\boldsymbol{Q_{\rm{S}}}}{T}\right).
\end{equation}

\subsection{Summary}
\label{subsec:conversion_summary}
To reformulate a total-energy-based HLLD Riemann solver, we need to replace the total energy with entropy in both predictor and corrector steps, if the Godunov method is used. Here are the steps:
\begin {enumerate}
\item Replace thermal energy-based EOS (eq. \ref{eq:eos_eth}) with entropy-based EOS (eq. \ref{eq:basic_S2});
\item In the predictor step, it is sufficient to advect $S$ as a passive scalar to get the source term $\sigma$, e.g. for the right state:
\begin{equation}
    \sigma_S = \frac{\Delta t}{2}\left( -\boldsymbol{u}\frac{ \partial s}{\partial t} \right) \pm
    		 \frac{\Delta \xi}{2}\left( -\boldsymbol{u}\frac{ \partial s}{\partial \xi} \right),
\end{equation}
where $\xi=[x,y,z]$ is the spatial dimension;
\item Compute the Godunov flux for entropy instead of total energy;
\item Compute the Godunov flux for kinetic and magnetic energies (see Appendix \ref{app1:numerical_method});
\item Compute $\boldsymbol{Q_{\rm{S}}}$ using equation \ref{eq:q_s};
\item Compute the $S_{\rm{gen}} = max\left(0,\frac{\boldsymbol{Q_{\rm{S}}}}{T}\right)$;
\item Apply Godunov flux, heating $\frac{\boldsymbol{Q_{\rm{ext}}}}{T}$ and $S_{\rm{gen}}$ to get $S^{t+1}$.
\end{enumerate}
In the next section we show a number of tests we have put this new solver through.

\section{The numerical tests}
\label{sec:numerical_tests}
To check the validity of the HLLS solver, we conduct a series of experiments. They are in 1D, 2D and 3D, single precision, both hydrodynamic and MHD. We run 1D tests in all principal directions, 2D tests in all 3 planes, to make sure the solver is well balanced in all dimensions. We note that some tests are run intentionally on low resolution, as resolution-starved experiments may reveal problems that are inversely proportional to resolution.

\begin{figure}
\centering
    \includegraphics[width=0.85\columnwidth]{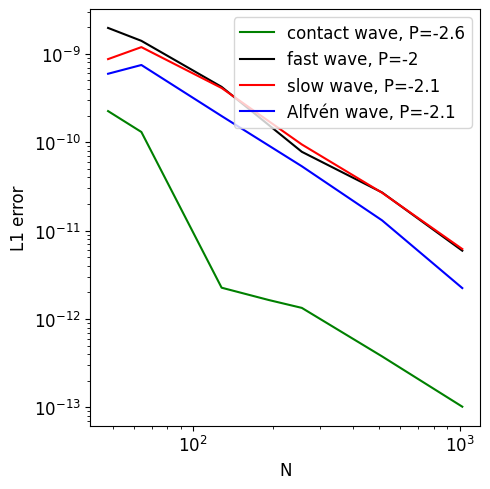}
    \caption{Convergence in the norm of the L1 error vector for contact, fast, slow, and Alfv\'en waves after propagating a distance of one wavelength. The convergence rate $P$ is given in the legend.}
    \label{fig:wave_convergence}
\end{figure}

\subsection{Linear wave convergence}
This test shows that the HLLS formulation converges with second order accuracy for linear amplitude waves. The test is nearly identical to section 8.2 in \cite{Stone_2008}, with a uniform medium where the generic quantities are $(\rho, \gamma,  P, \vb{u}, B_x, B_y, B_z) = [1,~ 5/3,~ 3/5,~ 0,~ 1,~ \sqrt{2},~ 1/2]$. The exact eigenfunctions for fast and slow magnetosonic, Alfv\'en, and contact waves for the background state are given in \cite{Gardiner_2005}. Since normally operate in single floating point precision, the original wave amplitude of $A= 10^{-6}$ is too close to the round-off error and is impractical to conduct an informative test. Thus, the only difference from \cite{Stone_2008} is that we set the wave amplitude value to be $A= 10^{-3}$. The resulting waves are still weak, do not steepen into shocks. Figure \ref{fig:wave_convergence} shows the norm of the $L_1$ error vector for each wave family as a function of the numerical resolution from 48 to 1024 cells. The figure shows that we achieve second-order accuracy in both space and time for smooth solutions.

\begin{figure}
\centering
    \includegraphics[width=0.85\columnwidth]{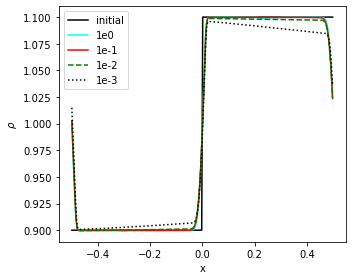}
    \caption{Entropy wave test. Density profiles at starting time (black) and for different wave speeds $u_x$: blue - 1/0, red - 0.1, green dashed - 0.01, magenta dotted - 0.001. The overlap of wave profiles is perfect, with expected slopes present.}
    \label{fig:entropy_wave}
\end{figure}

\subsection{Entropy wave}
\label{subsec:entropy_wave}
Riemann solvers are generally very good at preserving shocks. However, very slowly moving waves are sometimes more difficult. To test the diffusion and dispersion error, we launch a very slowly moving discontinuity (entropy wave) in one dimension. Assuming the experiment is carried out in $x$ direction, the generic quantities are $(\rho, P) = [0.9, 1.0]$ when $x<0$ and $[1.1, 1.0]$ when $x \geq 0$. All other quantities are set to 0, with $\gamma =\frac{5}{3}$ and $S= ln(\frac{P}{\rho^\gamma})$ in the entire experiment. Boundaries are periodic. We use rather low resolution, 100 cells in direction of the wave propagation to exaggerate the dispersion error. The velocity $u_x$ of the whole box is set constant to $[1.0, 0.1, 0.01, 0.001, 0.001]$ and the runtime of the experiment is set to respectively $t_{end}= [1, 10, 100, 1000, 1000]$ time units. This means the number of updates is increasing tenfold with each run. If the solver is diffusive, the wave form should smear out. Figure \ref{fig:entropy_wave} shows the test results at the end time, when the wave crosses one period. Naturally the corners are slightly rounded from slope limiters, but we can clearly see, that with lower velocities the increase in diffusion is negligible. The diffusion is resolution-sensitive, and depends on the slope limiters. The black curve is at time=0, blue - $u_x=1$, red - $u_x=0.1$, green dashed - $u_x=0.01$, and magenta dotted - $u_x=0.001$.
At the lowest velocity, we observe noticeable distortion. This is expected, as many Riemann solvers that are not asymptotic-preserving become excessively diffusive at low Mach numbers. The issue arises because numerical dissipation is often linked to jumps in the velocity normal to the cell interfaces, which do not scale correctly with Mach number. As a result, even small spurious variations in this velocity component can lead to disproportionately large diffusion, particularly when $\rm{M} \leq 0.2$.

\subsection{Shu \& Osher shocktube}
\label{subsec:shu_osher}
\begin{figure}
\centering
    \includegraphics[width=0.85\columnwidth]{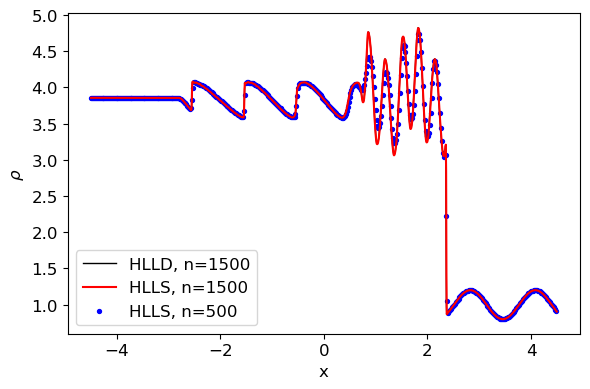}
    \caption{Shu \& Osher shocktube test. Density profile at time=1.8; black - reference HLLD solver with 1500 cells, red - HLLS solver with 1500 cells, blue dots - HLLS solver with 500 cells.}
    \label{fig:shu_osher}
\end{figure}

This test is a 1D Mach=3 shock interacting with sine waves in density \citep{Shu1989}. It tests the solver's ability to capture both shocks and small-scale smooth flow. The computational domain is 9 length units long and is split into two regions with different conditions in each of them. Assuming the experiment is carried out in $x$ direction, the quantities are $(\rho, u_x, P) = [3.857143, 2.629369, 10.33333]$ when $x<-4$ and $[1 +0.2 \sin{(5x)}, 0, 1]$ when $x \geq -4$. All the other quantities are set to 0, with $\gamma =1.4$ and $S= ln(\frac{P}{\rho^\gamma})$ in the whole experiment. Figure \ref{fig:shu_osher} shows the test results at time = 1.8 time units (red curve). The reference run is done with the HLLD solver (black curve, perfectly covered by the red curve). Both the test and the reference runs had 1500 cells. We additionally show a HLLS run with 500 cells (dots).

\subsection{Brio \& Wu shock tube}
\label{subsec:brio_wu}

\begin{figure}
\centering
    \includegraphics[width=1\columnwidth]{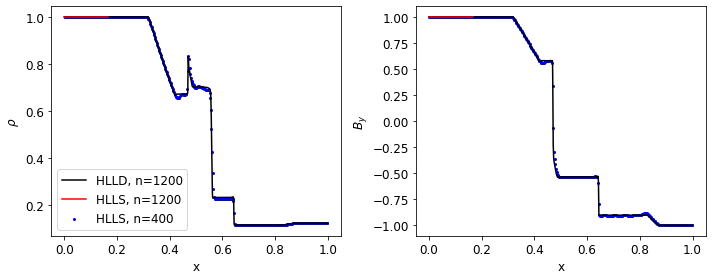}
    \caption{Brio \& Wu shock tube test. Density profile (left) and y component of magnetic field (right) at time=1.0; black - reference HLLD solver with 1200 cells, red - HLLS solver with 1200 cells, blue dots - HLLS solver with 400 cells. Black and red overlap each other nearly perfectly. }
    \label{fig:brio_wu}
\end{figure}

This is a classical test of an MHD shock tube, described by \citet[section V]{Brio1988400}, where the right and left states are initialised to different values. The left state is initialised as $(\rho, u_x, u_y, u_z, B_y, B_z, P) = [1,0,0,1,0,1]$, and the right state $[0.125,0,0,-1,0,0.1]$. $B_x=0.75$ and $\gamma = 2$. This test shows whether the solver can accurately represent the shocks, rarefaction waves, compound structures and contact discontinues. Figure \ref{fig:brio_wu} shows the test results at time=1.0. Black curve is for reference HLLD run with 1200 cells, red - HLLS run; blue dots - HLLS run with 400 cells. Both high and low resolution runs overlap each other nearly perfectly and follow the profiles in the literature very accurately. From left to right we can identify a fast rarefaction wave, compound wave, contact discontinuity, slow shock and a fast rarefaction wave again.

\subsection{Kelvin-Helmholtz instability}
\label{subsec:KH_instability}

\begin{figure}
\centering
    \includegraphics[width=0.425\columnwidth]{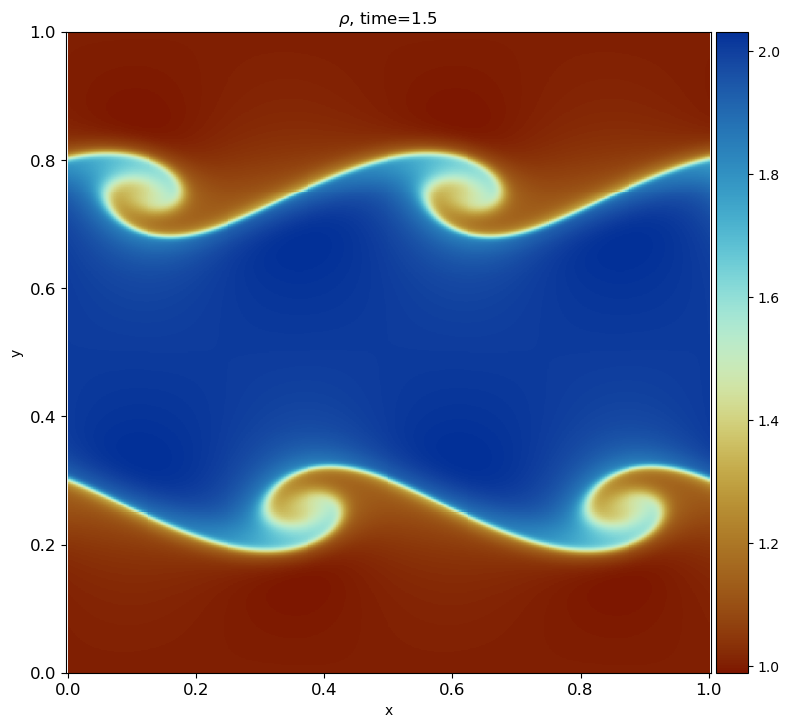}
    \includegraphics[width=0.425\columnwidth]{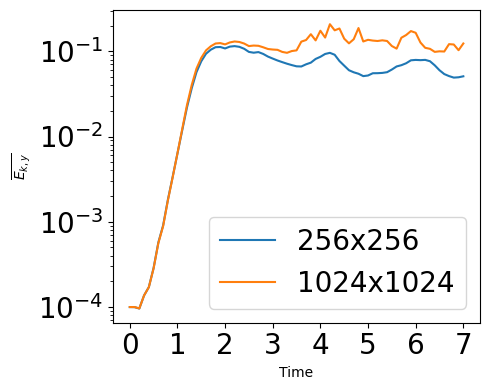}
    \caption{HD Kelvin-Helmholtz instability. Density profile (left) and time evolution of kinetic energy (right).}
    \label{fig:KH_instab_HD}
\end{figure}

   \begin{figure*}
   \resizebox{\hsize}{!}
   		 {\includegraphics[width=0.5\columnwidth]{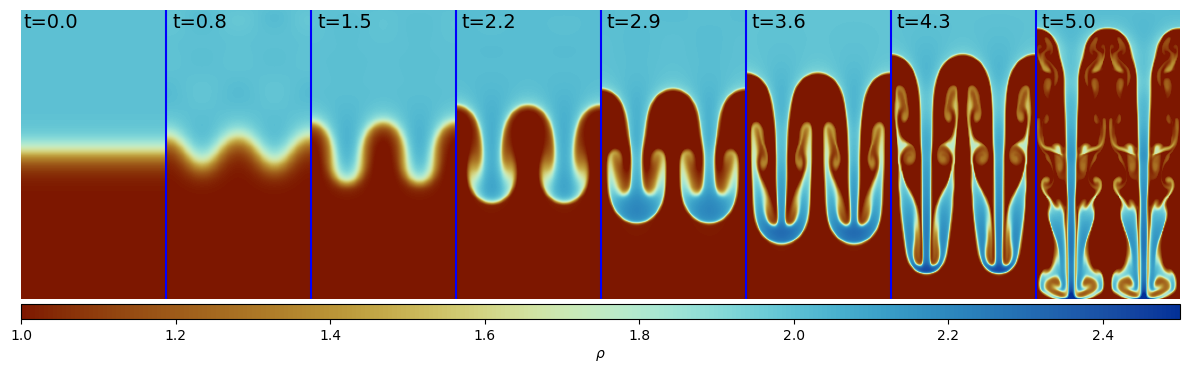}
      }
    \caption{Rayleigh-Taylor instability in low (128 $\times$ 256) resolution. We plot density at different times.}
    \label{fig:RT_instab}
   \end{figure*}

   \begin{figure*}
   \resizebox{\hsize}{!}
   		 {\includegraphics[width=0.5\columnwidth]{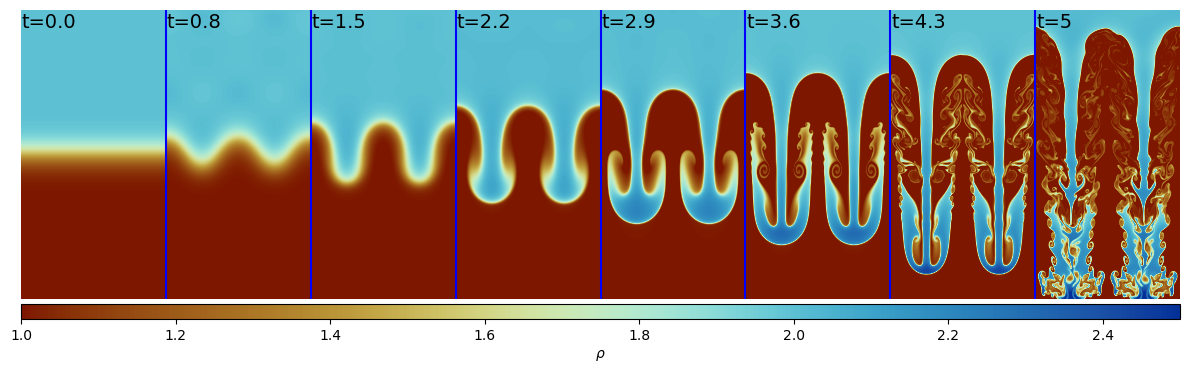}
      }
    \caption{Same as Fig. \ref{fig:RT_instab}, but in 768 $\times$ 1536 resolution.}
    \label{fig:RT_instab_HD}
   \end{figure*}
Kelvin-Helmholtz instability occurs when velocity shear is present within a continuous fluid or across fluid boundaries. We conduct the test as described in \cite{McNally_2012}. For convenience, here we summarise the setup. The domain is 1 unit by 1 unit in $x-$ and $y-$ directions, with resolutions of 256 $\times$ 256 cells. All boundaries are periodic. The density is given by
\begin{equation}
    \rho =
    \left\{
    \begin{array}{l}
    \rho_1 - \rho_m e^{\frac{y-1/4}{L}} \ \ \ \ \ \ \rm{if} \ 1/4 > y \geq 0    \\
    \rho_2 + \rho_m e^{\frac{1/4-y}{L}} \ \ \ \ \ \ \rm{if} \ 1/2 > y \geq 1/4  \\
    \rho_2 + \rho_m e^{-\frac{3/4-y}{L}} \ \ \ \ \rm{if} \ 3/4 > y \geq 1/2 \\
    \rho_1 - \rho_m e^{-\frac{y-3/4}{L}} \ \ \ \ \rm{if} \ 1 > y \geq 3/4   \\
    \end{array},
    \right.
\end{equation}
where $\rho_m = \left(  \rho_1 - \rho_2  \right)/2$, $\rho_1 = 1.0$, $\rho_2 = 2.0$ and $L=0.025$. The $x-$direction velocity is given by
\begin{equation}
    u_x =
    \left\{
    \begin{array}{l}
    u_1 - u_m e^{\frac{y-1/4}{L}} \ \ \ \ \ \ \rm{if} \ 1/4 > y \geq 0    \\
    u_2 + u_m e^{\frac{1/4-y}{L}} \ \ \ \ \ \ \rm{if} \ 1/2 > y \geq 1/4  \\
    u_2 + u_m e^{-\frac{3/4-y}{L}} \ \ \ \ \rm{if} \ 3/4 > y \geq 1/2 \\
    u_1 - u_m e^{-\frac{y-3/4}{L}} \ \ \ \ \rm{if} \ 1 > y \geq 3/4   \\
    \end{array},
    \right.
\end{equation}
where $u_m = \left( u_1 - u_2 \right) /2$, $u_1=0.5$ and $u_2=-0.5$. The background shear is perturbed by adding velocity in $y-$direction,
\begin{equation}
    u_y = 0.01 \sin{(4\pi x)}
\end{equation}
An ideal EOS with $\gamma=5/3$ is used. Initial gas pressure $P_{gas}=2.5$. We run the simulation until time $t = 10$. To do an exact comparison to codes in \cite{McNally_2012}, we show the gas density at $t=1.5$ and the maximal value of vertical kinetic energy evolution in time in figure \ref{fig:KH_instab_HD}. Note that we used 256x256 resolution, but the result is directly comparable to $512^2$ and $4096^2$ runs in \cite{McNally_2012}. On the right panel of the figure we show the maximum value of vertical kinetic energy in the simulation for three resolutions. From this figure it can be seen that primary instability sets on at exactly the same time, secondary instabilities occur around the same time as well, indicating that we get a converged solution even at the low resolution of 256x256.

\subsection{Rayleigh-Taylor instability}
\label{subsec:rt_instability}

Another classic test of a code’s ability to handle subsonic perturbations is the Rayleigh–Taylor instability and has been described in a number of studies, see e.g. \cite{Stone_2007,abel_2011,Hopkins_2015}. In this test a layer of heavier fluid is placed on top of a layer of lighter fluid. With gravitational source term added to the forces we test two things: whether the explicit addition of force is correct and the solver's ability to preserve instabilities while keeping things symmetric where it should be, during the linear phase. During the non-linear phase, in a sufficiently high resolution the symmetry is expected to break by construction. We use the initial setup similar to \citet{abel_2011,Hopkins_2015}. Here we recap the setup for convenience. In two dimensional domain with $0 \leq x \leq 0.5$ (128 cells for low and 768 for high resolution runs) and $0 \leq y \leq 1$ (256 and 1536 cells for low/high resolution runs respectively); we use periodic boundary conditions in $x$ direction and constant boundary conditions in $y$. In this test we use $\gamma=1.4$ and density profile is initialized as $\rho(y) = \rho_1 + (\rho_2 - \rho_1)/(1 + e^{-(y-0.5)/\Delta})$, where $\rho_1 = 1$ and $\rho_2 = 2$ are the density below and above the contact discontinuity ($y_c$ respectively, with $\Delta = 0.025$ being its width. The pressure gradient is in hydrostatic equilibrium with a uniform gravitational acceleration $g= -0.5$ in the $y$ direction, $P = \rho_2 / \gamma + g ~\rho(y) ~ (y - y_c)$, then entropy $S = \ln(P) - \ln(\rho)^\gamma$. An initial y-velocity perturbation $v_y = \delta v_y (1+\cos(8\pi(x+0.25)))(1+\cos(5\pi(y-0.5)))$ with $\delta v_y = 0.025$ is applied in the range $0.3\leq y \leq 0.7$ \citep{Hopkins_2015}.

Figure \ref{fig:RT_instab} shows the evolution of the instability at different times. The initial velocity grows and the heavier fluid starts to sink. Note the single-cell resolution of contact discontinuities and mixing. Both blobs are perfectly symmetric during the linear phase. Soon enough Kelvin-Helmholtz instabilities develop at the shear surface between the fluids, and the symmetry is harder to maintain due to low relative velocities. However, the low resolution run maintains perfect symmetry throughout the whole simulation. Figure \ref{fig:RT_instab_HD} shows the high resolution run. Secondary instabilities are more pronounced and thus the symmetry is harder to maintain. The figure can be directly compared to figure 22 in \citep{Hopkins_2015}. Our high resolution (768 $\times$ 1536) run shows very similar features as in \citep{Hopkins_2015}, although we maintain much better symmetries until the blobs reach the bottom.

\subsection{''Hot Bubble`` experiment}
\label{subsec:hot_bubble}

\begin{figure}
\centering
    \includegraphics[width=0.425\columnwidth]{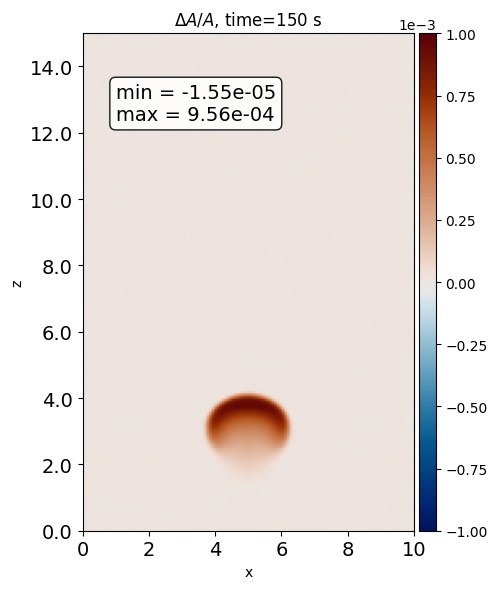}
    \includegraphics[width=0.425\columnwidth]{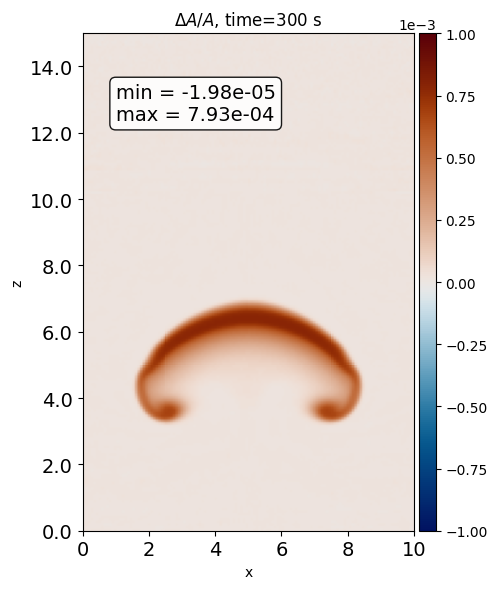}
    \caption{``Hot Bubble'' experiment at time = 150 and 300 s. Both panels show fluctuations of $\Delta A/A$, with the min/max values given in the floating box.}
    \label{fig:hot_bubble}
\end{figure}

To evaluate the entropy-preserving properties of our numerical scheme in a strongly stratified, low-Mach-number regime, we perform the "hot bubble" test following the setup of \cite{Edelmann_2021}. This experiment simulates the buoyant rise of a localized entropy perturbation embedded in an isentropic but highly stratified layer. In stellar interiors, entropy fluctuations drive convection, but outside the heating or cooling layers, these fluctuations should remain conserved except for mixing effects. Maintaining accurate entropy advection in such conditions is usually numerically challenging due to the large variations in pressure and density. The experimental setup is identical to the original, with a domain resolution of 128 $\times$ 192 cells in $[x,z]$ domain is used. The gravitational acceleration is then
\begin{equation}
    g_z = g_0 \sin(k_z z),
\end{equation}
where $g_0 = -1.099044 \times 10^5$ cm and $k_z = 2\pi /z_{max}$, $z_{max}=1.5 \times 10^6$ cm, while $x_{max}$ is $1 \times 10^6$ cm. $\gamma = 5/3$, at $z=0$ $P_0 = 10^6$ Ba, with temperature $T_0 = 300$ K. The stratification is isentropic, i.e.
\begin{equation}
    \rho(z) = \left(\frac{P(z)}{A} \right)^{1/\gamma},
\end{equation}
where $A=A_0=const$ everywhere outside of the bubble. As the $A_0$ value is not given in \cite{Edelmann_2021}, we use the ideal gas EOS to get $\rho_0 = P / R_{gas}T = 4.00906 \times 10^{-5}$ and consequently $A0 = 2.129425 \times 10^{13}$. $A=P/\rho^\gamma$ is not a true entropy expression, as discussed in section \ref{sec:S_EOS}, but a rather good probe for it. Since in ideal gas entropy can be simplified to $S=ln(P/\rho^\gamma)/(\gamma-1)$, we can express $S=S_0=ln(A)/(\gamma-1)=46.03419$. The hydrostatic pressure stratification is given by
\begin{equation}
    P(z) = \left[ P_0^{1-\frac{1}{\gamma}}+ \left( 1-\frac{1}{\gamma} \right) \frac{g_0}{A_0^\frac{1}{\gamma}k_z} \left[1 - \cos(k_z z) \right] \right]^{\frac{\gamma}{\gamma-1}},
\end{equation}
which is not perturbed and has a low-to-high ratio of 100. The perturbation is applied only for density via
\begin{equation}
    A = A_0 \left[ 1+ \left( \frac{\Delta A}{A} \right)_{t=0} \cos\left( \frac{\pi r}{2 r_0} \right)^2 \right],
\end{equation}
where $(\Delta A /A)_{t=0} = 10^{-3}$ is the bubble's amplitude and $r = \left[ \left(x-x_0 \right)^2 + \left( y-y_0 \right)^2 \right]$ is the distance from the bubble's centre. The radius of the bubble is $r_0=1.25 \times 10^5$ cm. This low perturbation makes the bubble rise at low Mach numbers, a few times $10^{-2}$ \citep{Edelmann_2021}. To improve the numerical precision we rescale the length scale, $l_{sc} = 10^{5}$, and density scale, $d_{sc} = 4 \times 10^{-5}$ leaving the time scale intact. This results velocities, density, entropy, and entropy being closer to unity, which in turn leads to highest numerical accuracy. And since we will look at the normalized fluctuations in A, $\Delta A / A$, the scaling does not change the result. The simulation is run from time $t=0$ to $t=300$ s. Two time slices at t=150 and 300 s are shown in Figure \ref{fig:hot_bubble}. This figure can be directly compared to Fig. 6 in \cite{Edelmann_2021}. The HLLS solver performs almost as good as a well-balanced scheme and noticeably better than simulations with energy as the primary thermodynamic variable without any well-balancing schemes. We note, that the main challenge for us is the perturbation being very close to the numerical noise limit as we normally operate with single precision. The adherence to the second law of thermodynamics is ensured to the numerical precision via eq. \ref{eq:S_update}, thus the negative entropy, and consequently $A$, values are from numerical round-off errors.
\subsection{MHD blast}
\label{subsec:blast}

\begin{figure}
\centering
    \includegraphics[width=0.425\columnwidth]{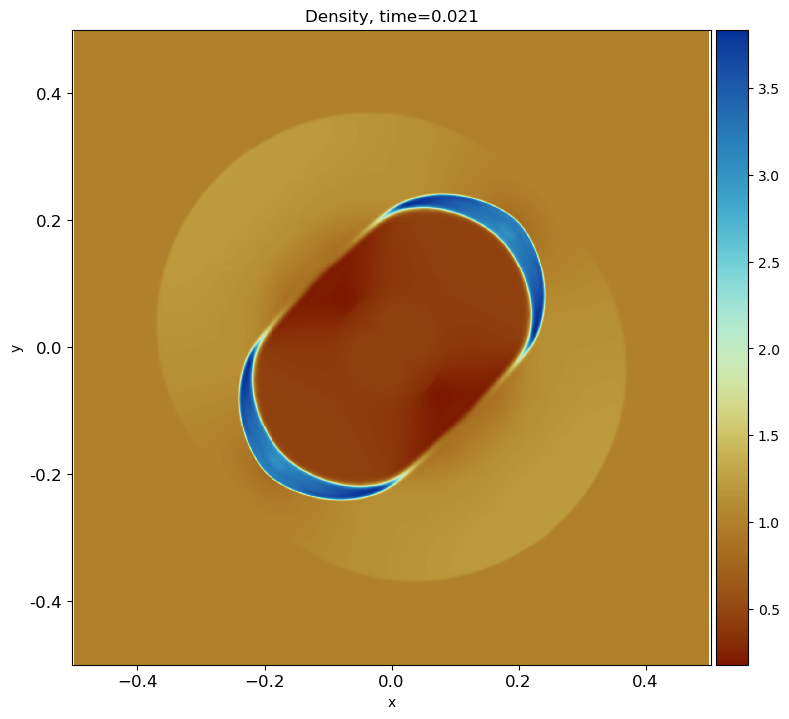}
    \includegraphics[width=0.425\columnwidth]{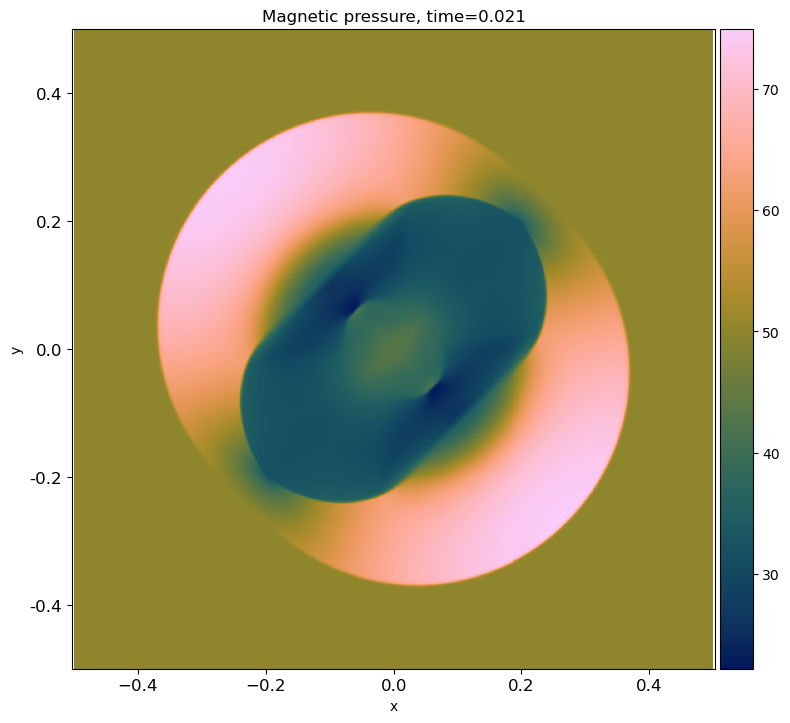}
    \caption{MHD blast experiment. Density (left) and magnetic pressure (right) at time=0.021. Note the sharp shock-fronts.}
    \label{fig:blast}
\end{figure}

\begin{figure*}
  \centering
  \includegraphics[width=0.65\columnwidth]{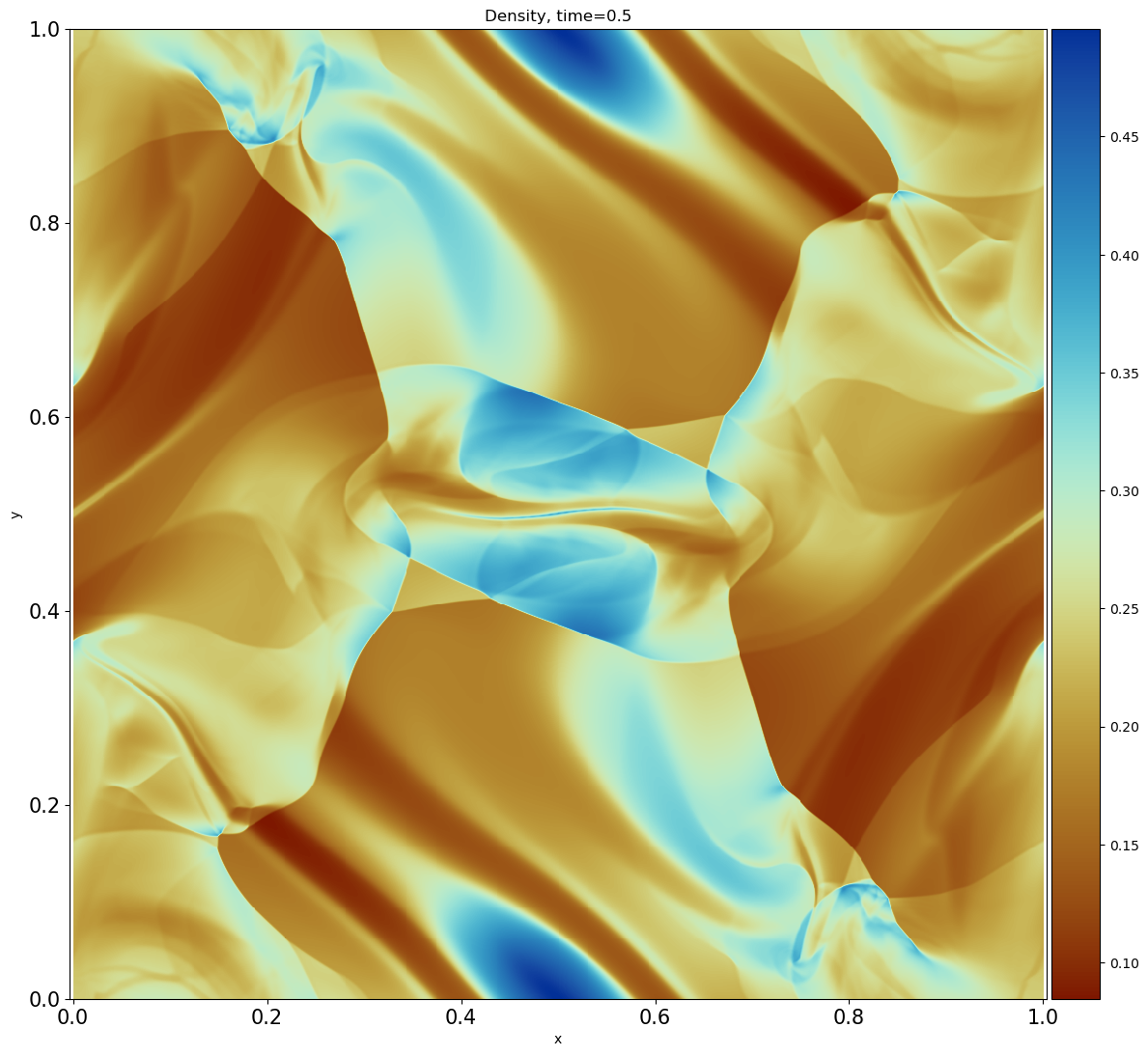}
  \includegraphics[width=0.65\columnwidth]{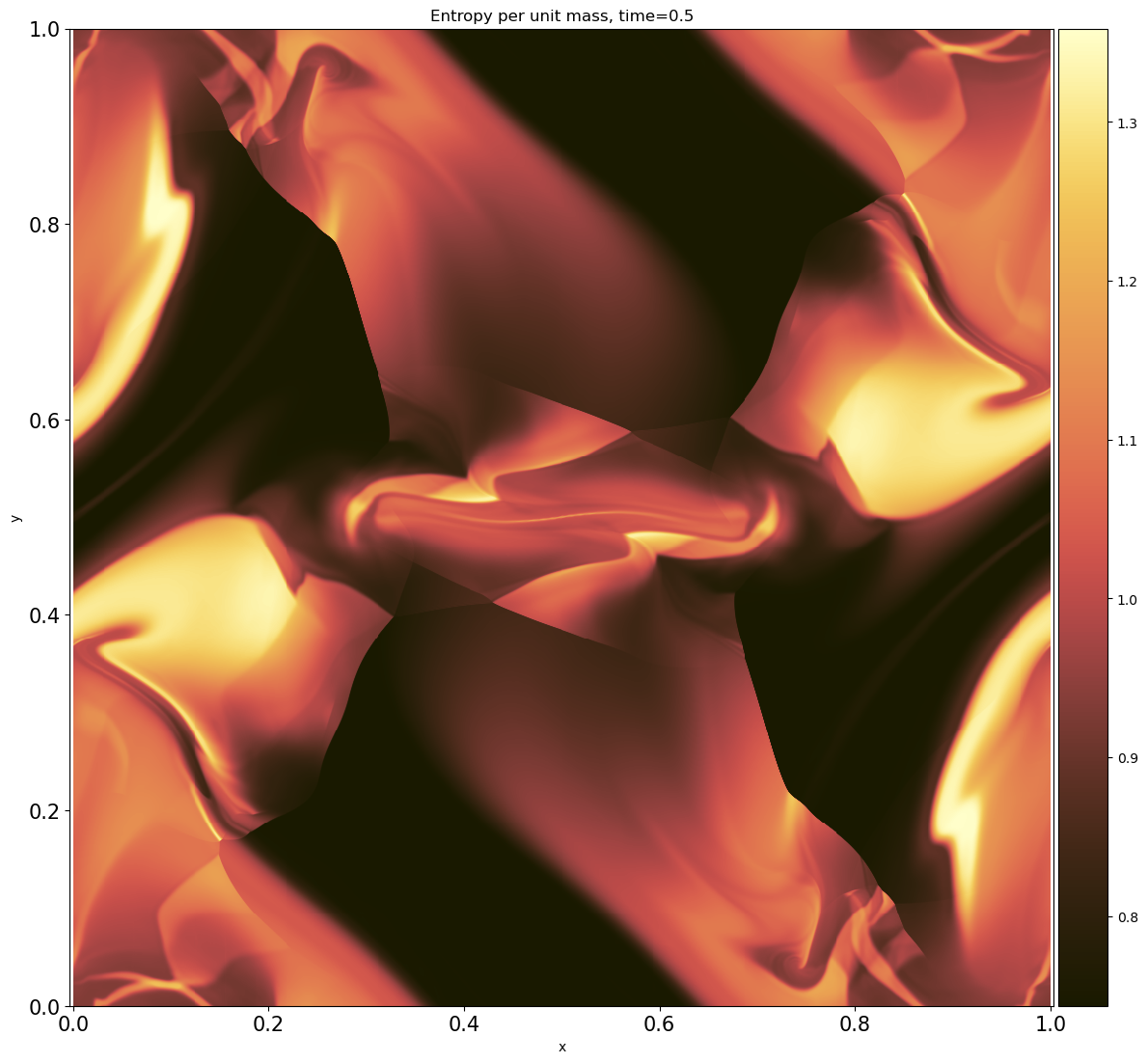}
  \includegraphics[width=0.65\columnwidth]{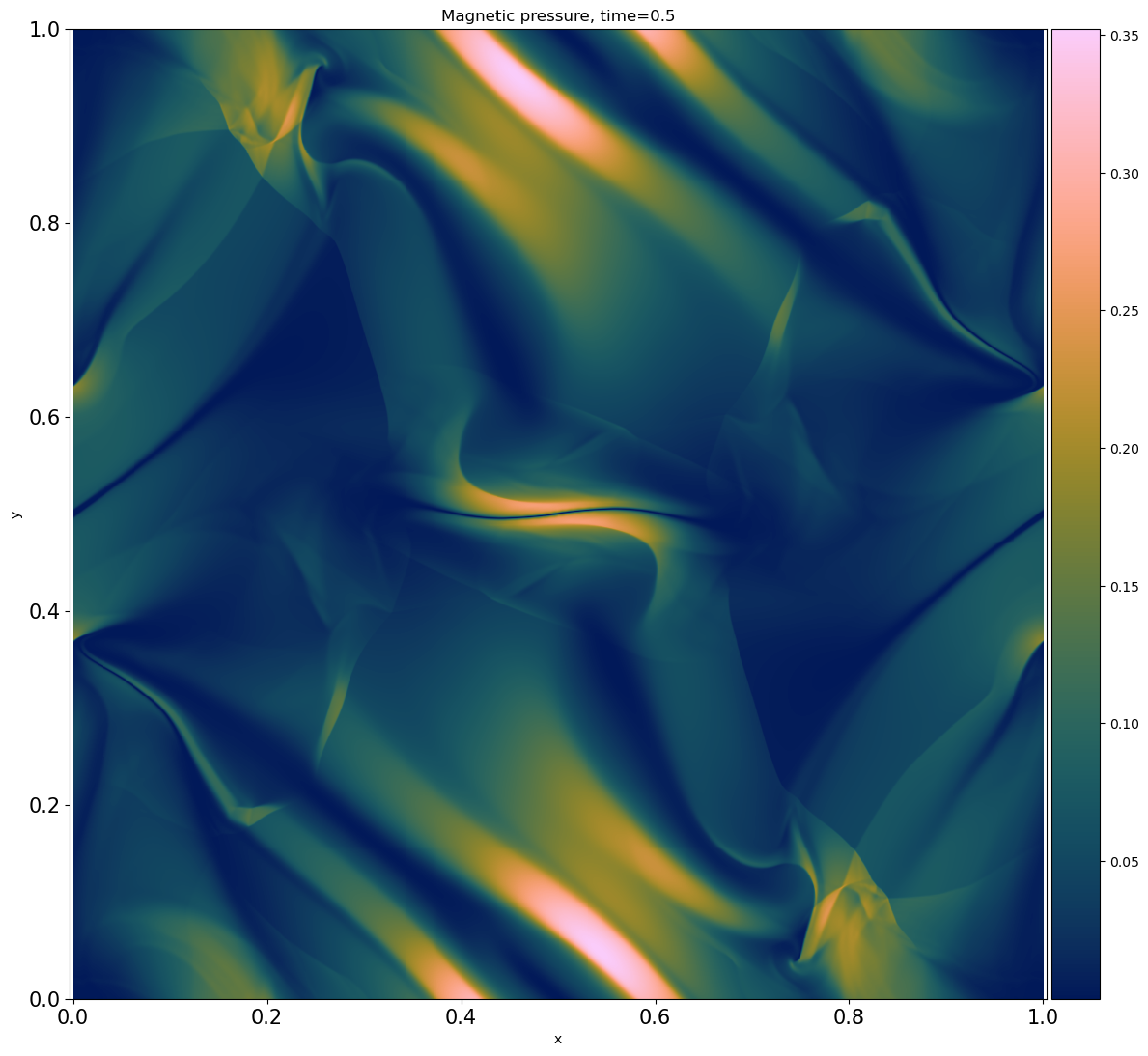}
  \caption{Orszag-Tang vortex test at time = 0.5. Density (left), entropy per unit mass (middle) and magnetic pressure (right).}
  \label{fig:OT_vortex_05}
\end{figure*}

This is a very popular test and various papers have presented results with slightly different problem setups. It is a very good test of the code's ability to handle the evolution of strong MHD waves and look for directional biases. We chose to follow the setup by e.g. \cite{Ramsey_2012,Clarke_2010,Stone_2008}. A rectangular domain with $-0.5 \leq [x,y] \leq 0.5$, 512 $\times$ 512 cells resolution is used. All boundaries are periodic. $(\rho, u, B_x, B_y, B_z) = (1,0,5\sqrt{2},5\sqrt{2},0)$. The ambient pressure $P_{gas} =1$ with $P_{gas}=100$ within radius $r = 0.125$. Figure \ref{fig:blast} shows the density and magnetic pressure in the experiment at time $t=0.021$ and can be directly compared to \cite{Ramsey_2012,Clarke_2010,Stone_2008}. The shock front is sharp, a fast magneto-acoustic wave is travelling at the correct speed, the blast is symmetric.
\subsection{Orszag-Tang vortex}
\label{subsec:OT_vortex}

\begin{figure*}
  \centering
  \includegraphics[width=0.65\columnwidth]{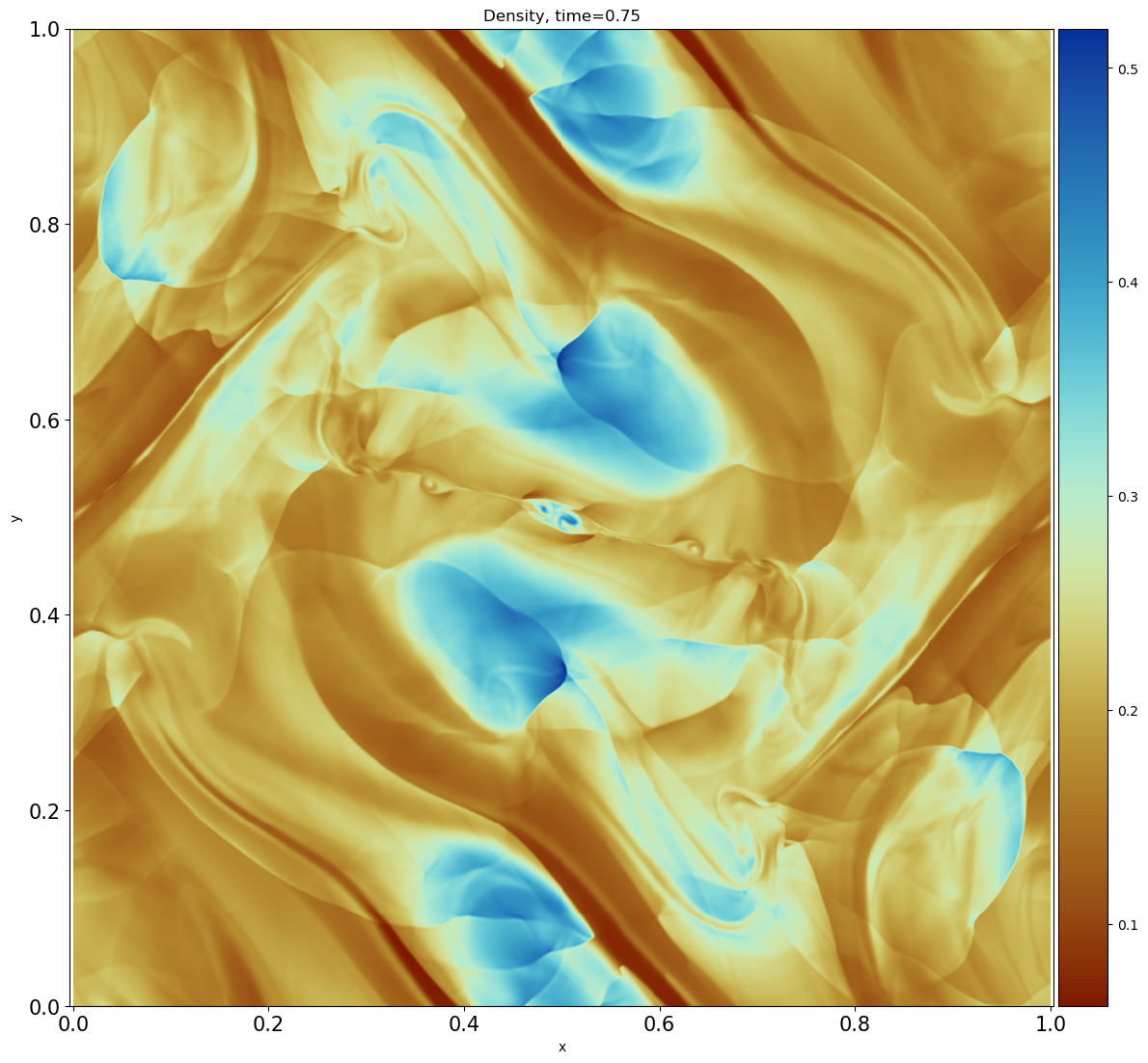}
  \includegraphics[width=0.65\columnwidth]{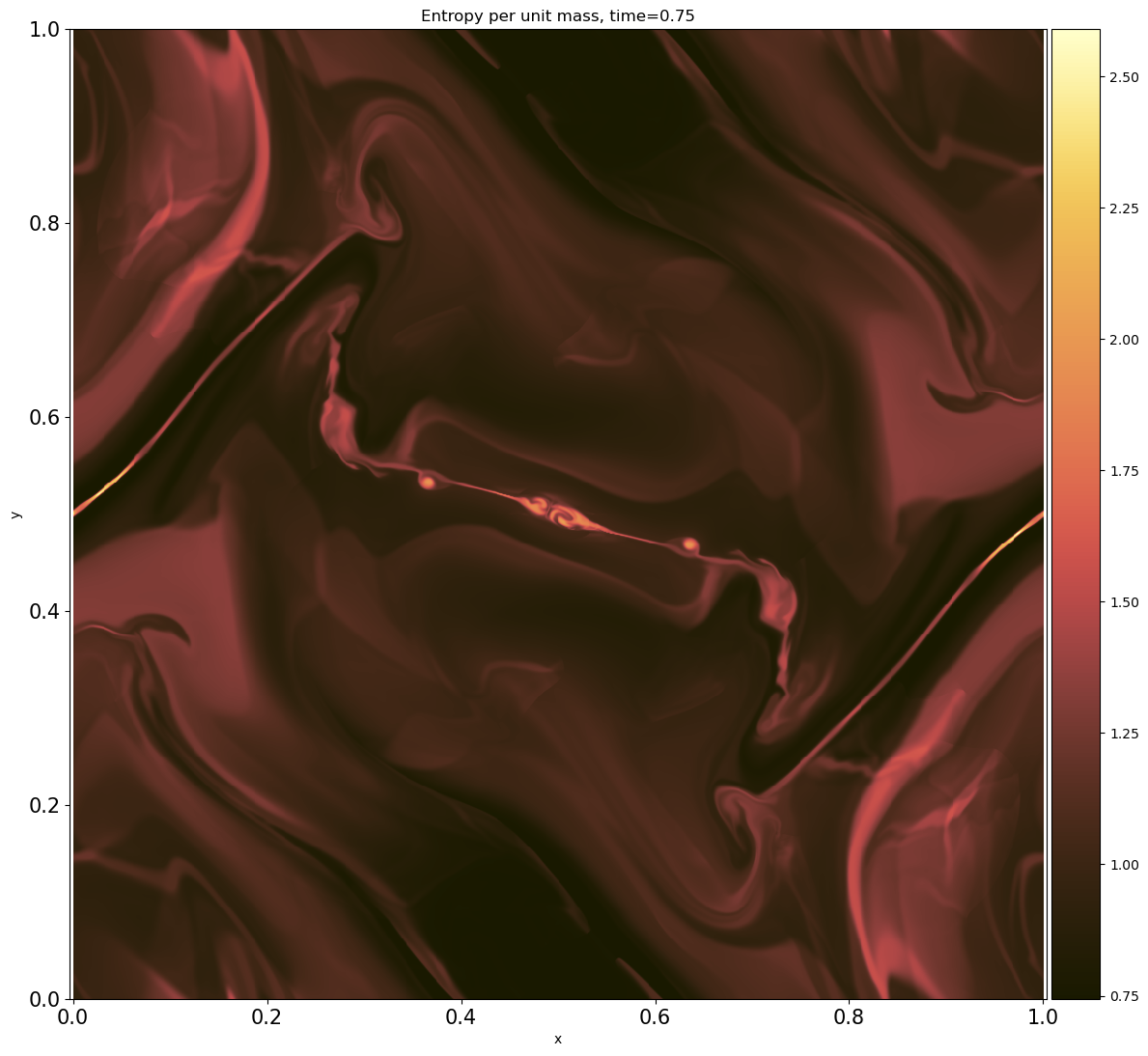}
  \includegraphics[width=0.65\columnwidth]{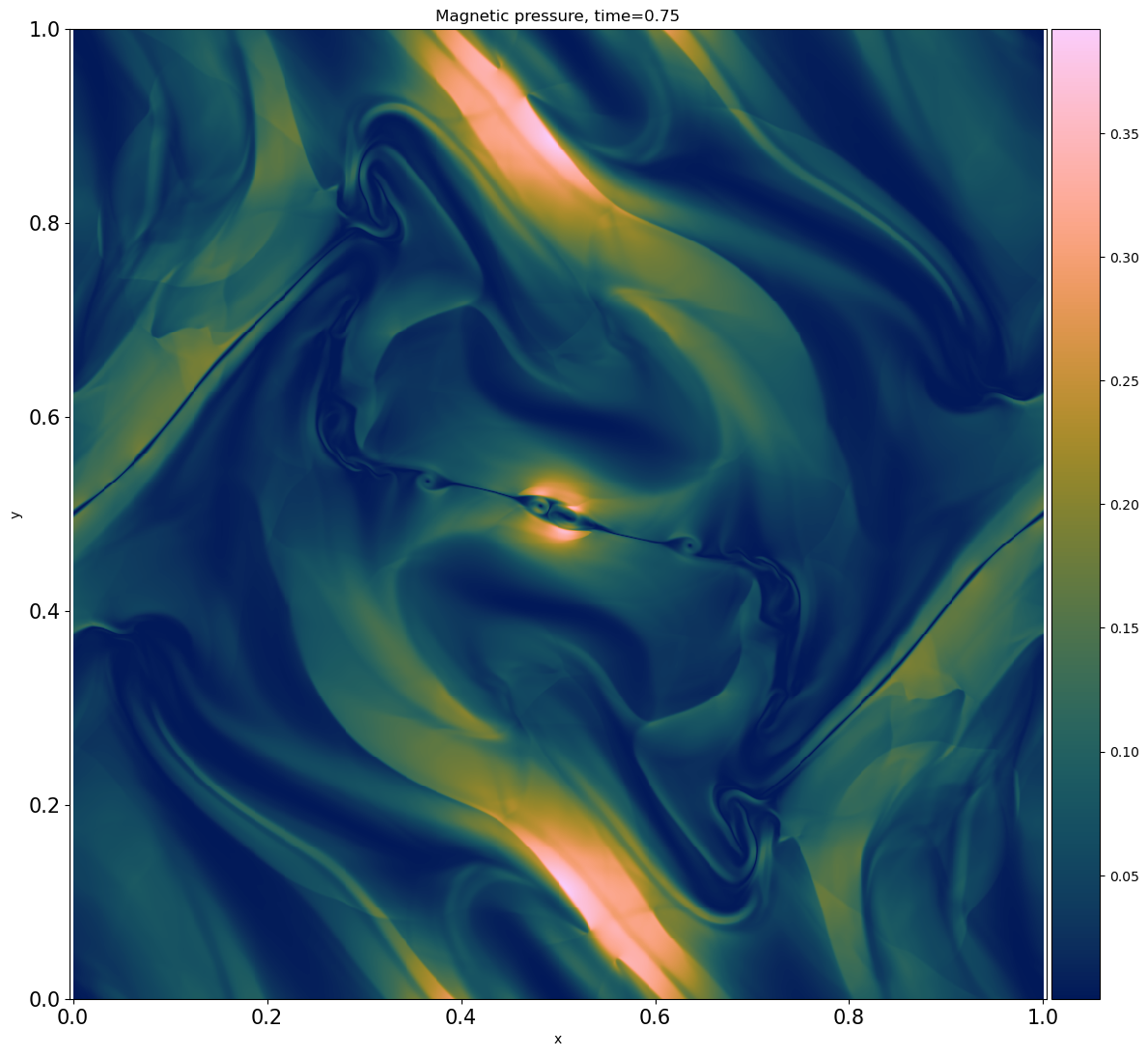}
  \caption{Orszag-Tang vortex test at time = 0.75. Density (left), entropy per unit mass (middle) and magnetic pressure (right). Notice the plasmoids forming.}
  \label{fig:OT_vortex_075}
\end{figure*}

This test has become a staple for MHD codes. The initial setup is identical to \cite{Stone_2008,Ramsey_2012}. A rectangular domain with $-0.5 \leq [x,y] \leq 0.5$, 1024 $\times$ 1024 cells for high resolution run is used. All boundaries are periodic. Initially the pressure and density are constant, $P_{gas}=5/12\pi$ and $\rho=25/36\pi$. The ratio of specific heats $\gamma = 5/3$. Initial velocity $(u_x, u_y, u_z) = (-\sin{(2\pi y)}, \sin{(2\pi x)},0)$ and the magnetic field is set through the vector potential
\begin{equation*}
A_z = \frac{B_0}{4\pi}\cos{(4\pi x)}+\frac{B_0}{2\pi}\cos{(4\pi y)},
\end{equation*}
where $B_0 = 1/\sqrt{4\pi}$. Figures \ref{fig:OT_vortex_05} and \ref{fig:OT_vortex_075} shows density, entropy per unit mass and magnetic pressure at two times, $t=0.5$ and $t=0.75$. The first time is the typical time this test is shown. When compared to \cite{Stone_2008,Ramsey_2012}, we can recognize similar features. Notice the very sharp features and perfect symmetry between sides. When advancing the simulation further, the vortex starts producing plasmoids. Here it is very difficult to maintain symmetry, but the lower panels show that plasmoids are released rather symmetrically.

   \begin{figure*}
  \centering
  \includegraphics[width=1.99\columnwidth]{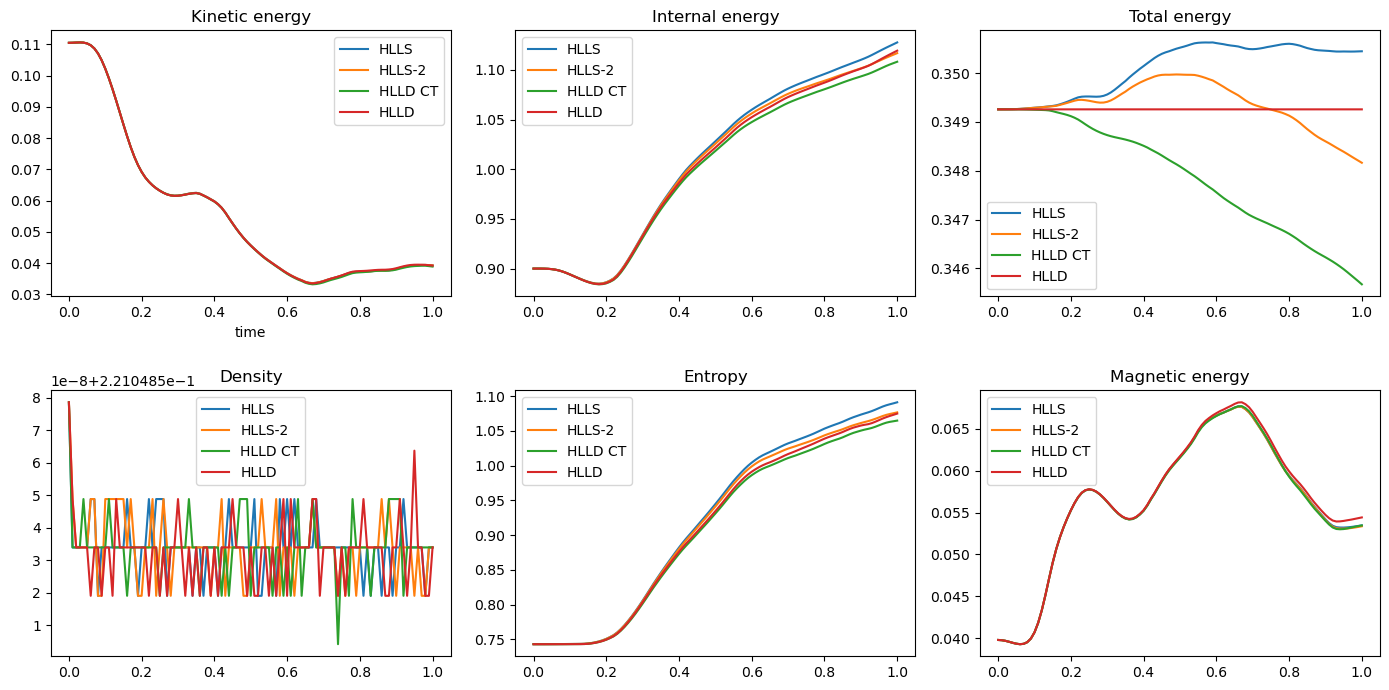}
  \caption{Current-sheet test at different times. Density (top) and magnetic pressure (bottom).}
  \label{fig:OT_time_series}
\end{figure*}

This experiment is very effective testing the solver stability. In high resolution, discontinuities and rarefactions are more severe and Riemann solvers can crash from negative pressure or thermal energy values, if they cannot deal with them. Since we have a HLLD solver, that can use both total and thermal energies as a primary thermodynamic variable available as well, we can investigate the conservation of different quantities in different formulations. Figure \ref{fig:OT_time_series} shows the time evolution of horizontally averaged kinetic (top left), magnetic (bottom right), internal (top middle) and total (top middle) energies, as well as density (bottom left) and entropy (bottom middle). The resolution was identical in all the runs (256x256). HLLS represents the run with the solver, presented in this paper, HLLS-2 allows $\boldsymbol{Q_{S}}$ to be negative, breaking the second law of thermodynamics, but more accurately following the energy dissipation and numerical dispersion (see discussion in subsection \ref{subsubsec:negative_qs}); HLLD is a classical HLLD solver where total energy is conserved, magnetic fields are updated using the CT method, but no adjustments to the total energy are done; while HLLD CT is a modification of the said solver, where the magnetic energy contribution in the total energy is updated with the values from CT solver, making it more stable in areas, where magnetic energy strongly dominates over kinetic energy. This correction makes the solver to not conserve the total energy in favour of conserving all three energies separately. 

In the figure we can see, that HLLS solver does not conserve the total energy, by construction. However, if we allow $\boldsymbol{Q_{S}} < 0$, it tracks the total energy conserving HLLD solver rather close, indicating where the second law of thermodynamics is broken in HLLD solver. The prime reason for this non-conservation is the discrepancy of magnetic fields between the 1D Riemann solver and the CT 2D solver. The former is more diffusive, assumes normal magnetic field is constant across the interface, while the latter is much less diffusive and preserves $ \div{\vb{B}} = 0$ better. Secondary effects stem from numerical dispersion. 

However, the discrepancy between the solvers diminishes with higher resolution. In realistic simulations open boundary conditions, external forces and higher desired resolution may deem the discrepancies in total energy negligible. 

\subsection{Current sheet}
\label{subsec:current_sheet}

   \begin{figure*}
  \centering
  \includegraphics[width=1.99\columnwidth]{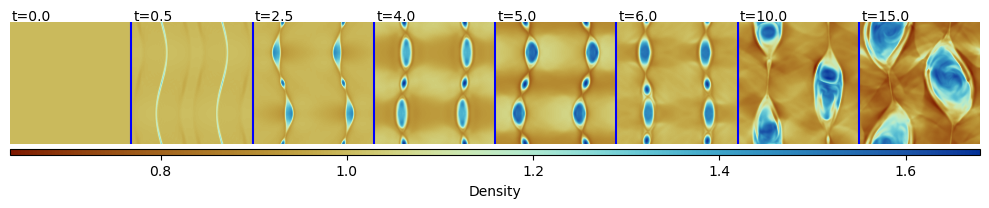}
\\
  \includegraphics[width=1.99\columnwidth]{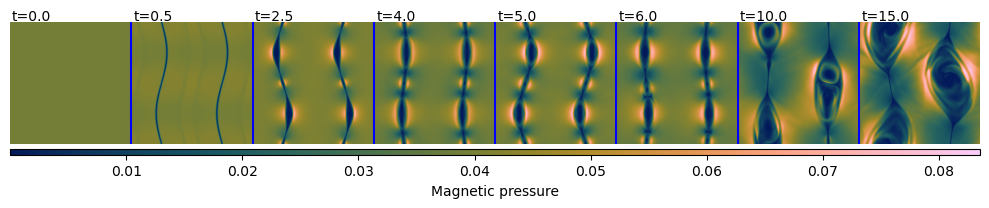}
  \caption{Current-sheet test at different times. Density (top) and magnetic pressure (bottom).}
  \label{fig:current_sheet}
\end{figure*}

The current sheet test problem is designed to see what an algorithm will do with a perturbed current sheet. Although an analytic solution for this test problem is not available, it is still suitable for test the robustness of the algorithm. The experimental setup is similar to \cite{Hawley_1995}. The test is run in two dimensions using a periodic, square grid  with $-0.5 \leq [x,y] \leq 0.5$, 256 $\times$ 256 cells resolution. There is a uniform magnetic field that discontinuously reverses direction at a certain point. In the whole box $(\rho, P_{gas}) = (1,\beta/2$, where $\beta$ is an input parameter. We set $B_y$:
\begin{equation}
    B_y =
    \left\{
    \begin{array}{l}
    1/\sqrt{(4\pi)} \ \ \ \ \ \ \ \rm{if} \ \left|x\right| > 0.25    \\
    -1/\sqrt{(4\pi)} \ \  \ \ \rm{if} \ \left|x\right| \leq 0.25    \\
    \end{array},
    \right.
\end{equation}

For $\left| x \right| > 0.25$, velocities $(u_x,u_y) = (A \sin{(2\pi y)},0)$, where $A$ is an amplitude. We ran the test with a range of $\beta$ and $A$ values, but here we present only with the "standard" setup, where $\beta = A = 0.1$. In figure \ref{fig:current_sheet} we can see the density and magnetic pressure at different times.
Initially linearly polarised Alfv\'en waves propagate along the field in the y-direction and quickly start generating magneto-acoustic waves since the magnetic pressure does not remain constant. Since there are two current sheets in the setup (at $x= \pm 0.25$, reconnection inevitably occurs. If $\beta < 1$, this reconnection drives strong over-pressurized regions that launch magneto-acoustic waves transverse to the field. Moreover, as reconnection changes the topology of the field lines, magnetic islands form, grow, and merge. The point of the test is to make sure the algorithm can follow this evolution for as long as possible without crashing. We ran tests for $0.1 \leq (\beta, A) \leq 10$ and the HLLS solver did not encounter any numerical problems.

\subsection{Magnetic field loop advection}
\label{subsec:mag_loop_adv}

\begin{figure}
\centering
    \includegraphics[width=0.99\columnwidth]{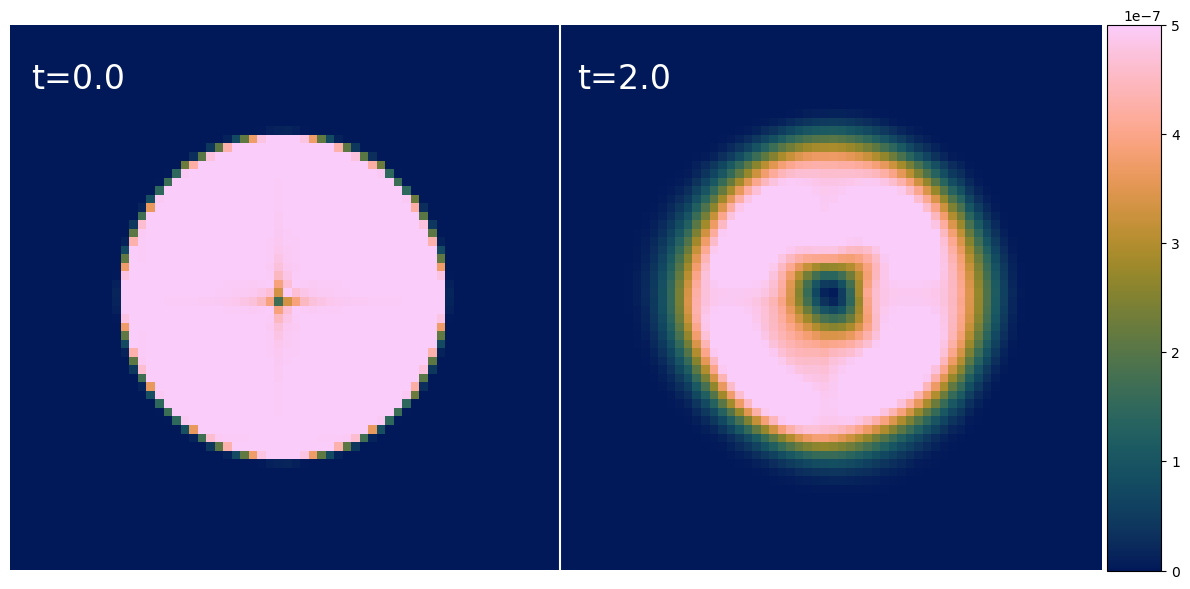}
    \caption{Magnetic loop advection experiment. The initial magnetic pressure (left) and after the loop has been advected around the grid twice.}
    \label{fig:mag_loop}
\end{figure}

This is a very powerful test to check whether the scheme preserves $\nabla \cdot B = 0$.  The experiment is similar to \cite{Toth_1996,Stone_2008} with two dimensional domain, $0 \leq x \leq 2$ (128 cells) and $0 \leq y \leq 1$ (64 cells). Boundaries are periodic in the whole domain. In this test $\gamma=5/3$ and both density, and gas pressure are constant throughout the box, $\rho = P_{\rm{gas}} = 1.0$. The magnetic field is initialised using a vector potential $A_z = \rm{max}\left[ A*(r_0-r),0 \right]$, with $A=10^{-3}$, $r_0=0.3$ and $r$ is the radial distance from the domain centre. The flow velocity $(u_x,u_y)=(2,1)$, thus the problem is essentially an advection test for the vector potential.

Although this test does not test the HLLS solver directly, the CT scheme nevertheless is an integral part of the solver. Moreover, we found this test to be a particularly useful tool to find errors in the magnetic fields predictor step. Figure \ref{fig:mag_loop} shows the initial magnetic pressure and after the magnetic loop has been advected twice through the domain. The shape of the loop is rather well preserved.

   \begin{figure*}
  \centering
  \includegraphics[width=0.4\columnwidth]{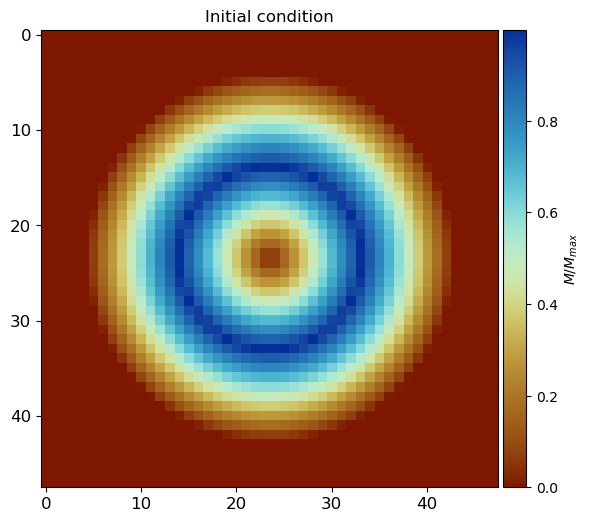}
  \includegraphics[width=0.4\columnwidth]{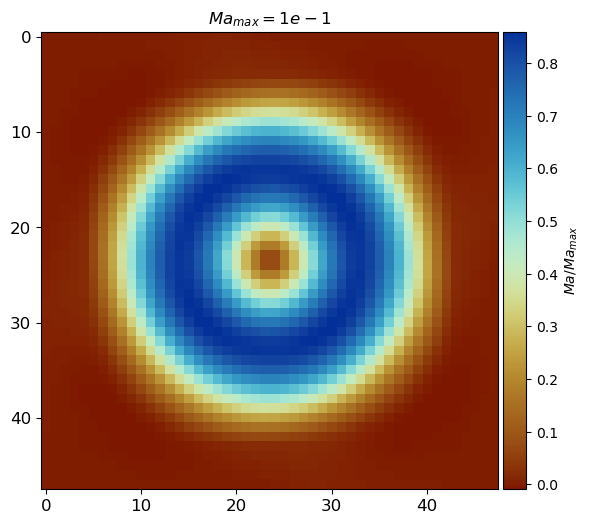}
  \includegraphics[width=0.4\columnwidth]{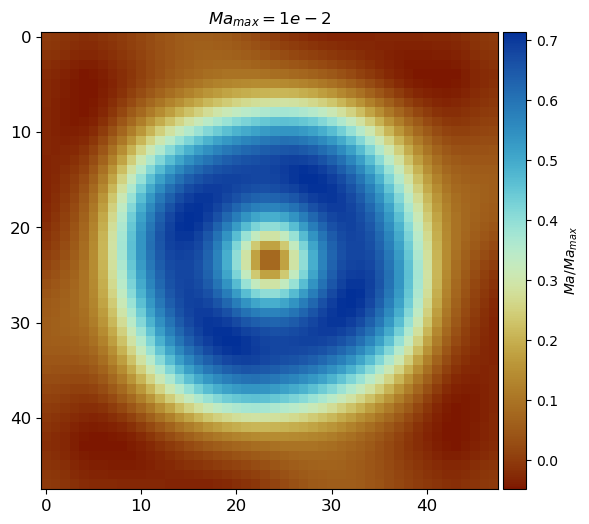}
  \includegraphics[width=0.4\columnwidth]{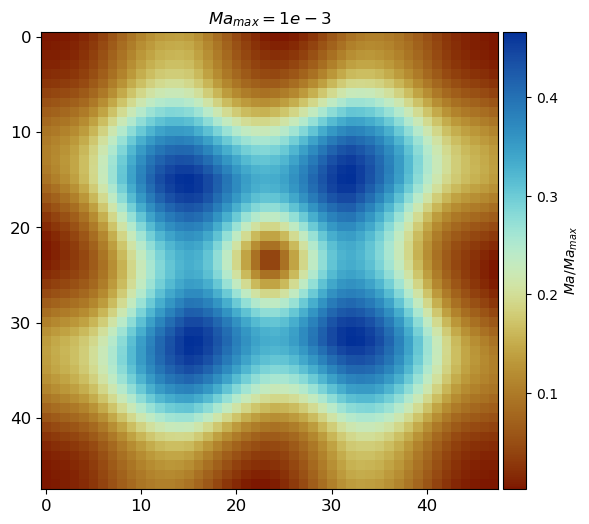}
  \includegraphics[width=0.4\columnwidth]{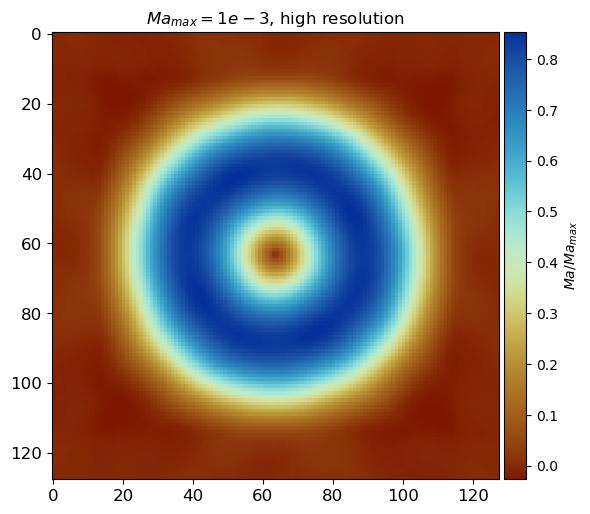}
  \caption{Gresho vortex test. The initial setup (left) and after 2 rotation periods with different Mach numbers, $Ma= 0.1, 0.01, 0.001$. The rightmost panel is for a higher resolution run. See text for more details.}
  \label{fig:gresho}
\end{figure*}

\begin{figure}
\centering
    \includegraphics[width=0.425\columnwidth]{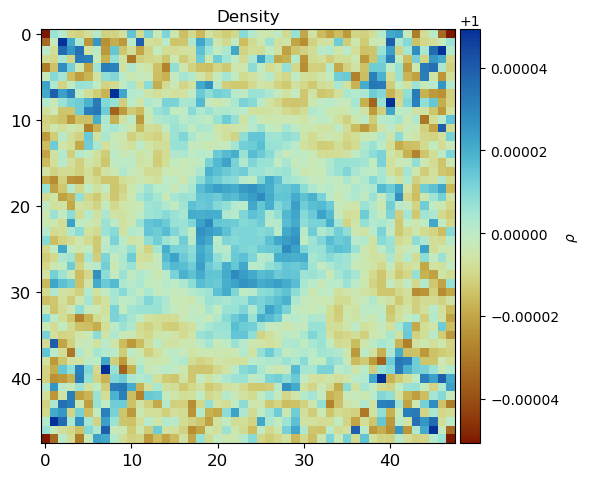}
    \includegraphics[width=0.425\columnwidth]{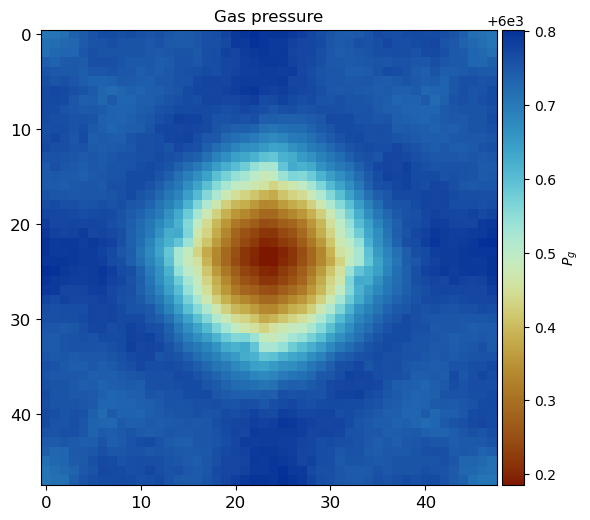}
    \caption{Gas density (left) and pressure (right) in the Gresho vortex test with $Ma=0.01$. Note, that the quantities are within the noise limit.}
    \label{fig:gresho2}
\end{figure}

\subsection{Gresho vortex}
\label{subsec:gresho_vortex}

The Gresho vortex is a time-independent rotation pattern. Angular velocity depends only on the radius and centrifugal force is balanced by the pressure gradient. We use the slightly modified initial condition, which permits the variation of the Mach number, it can be found in \cite{Happenhofer_2013,Grimm_2014}. For convenience, the setup is summarised here. The simulation is in two dimensions, $0 \leq x \leq 1$ (48 cells) and $0 \leq y \leq 1$ (48 cells) with periodic boundary conditions everywhere. The low resolution is deliberate - from our tests we see that with higher resolution (e.g. \citealt{Grimm_2014}) the experiment becomes much less challenging to the solver. In the whole domain $\rho = 1$, $\gamma=5/3$,
\begin{equation}
    u_\phi =
    \left\{
    \begin{array}{l}
    5r \quad \quad \ \rm{if} \ 0 \leq r < 0.2    \\
    2 - 5r \quad \rm{if} \ 0.2 \leq r \leq 0.4  \\
    0 \quad \quad \quad \rm{if} \ 0.4 < r \\
    \end{array},
    \right.
\end{equation}
\begin{equation}
    P_{gas} =
    \left\{
    \begin{array}{l}
    P_0 + \frac{25}{2}r^2 \quad \quad \ \rm{if} \ 0 \leq r < 0.2    \\
    P_0 + \frac{25}{2}r^2+ 4(1- 5r - \ln(0.2r) \ \rm{if} \ 0.2 \leq r \leq 0.4  \\
    P_0 - 2 +4\ln(2) \quad \quad \quad \rm{if} \ 0.4 < r \\
    \end{array},
    \right.
\end{equation}
where  $r=\sqrt{(x^2+y^2)}$ is the radial distance from the centre of the domain, $u_\phi$ is the angular velocity in terms of the polar angle $\phi=\rm{atan2}(y,x)$ and
\begin{equation}
    P_0 = \frac{\rho}{\gamma \rm{Ma}^2_{\rm{ref}}},
\end{equation}
with Ma$_{\rm{ref}}$ being a reference Mach number, which is the highest Mach number in the resulting flow. 4 runs are executed, with $Ma_{\rm{ref}}=[0.1, 0.01, 0.001]$. The last $Ma_{\rm{ref}}$ is repeated in two runs - the nominal resolution and a higher resolution, 128 $\times$ 128 cells. Figure \ref{fig:gresho} shows the results. We deliberately did not run the experiment with larger Mach numbers, as those are just too easy to maintain and the final result looks identical to the initial condition. In the figure we can see, that the vortex is maintained very well down to Ma=0.01, and with Ma=0.001 we get the expected outcome - the Godunov type Riemann solvers deal with very low Mach numbers poorly, even though Ma=0.01 result is still very good. Especially since we use single floating point precision. Figure \ref{fig:gresho2} shows the gas density and pressure with Ma=0.01. It can be clearly seen that the vortex itself is barely distinguishable inside the noise of the box.
On the other hand, with increased resolution, the result with Ma=0.001, as expected, has significantly improved and the vortex can be once again identified.

   \begin{figure*}
  \centering
  \includegraphics[width=0.5\columnwidth]{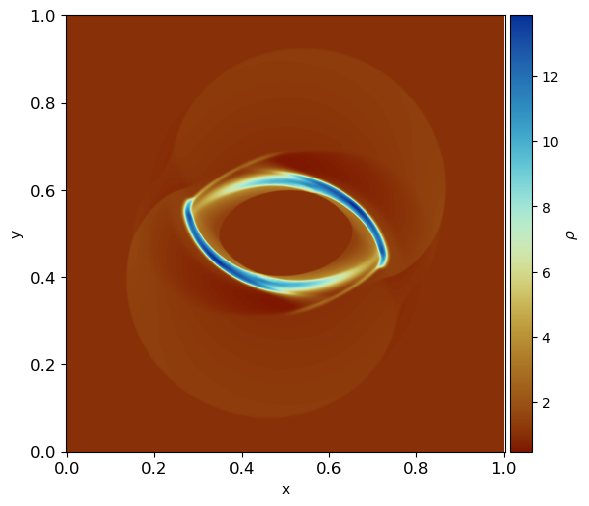}
  \includegraphics[width=0.5\columnwidth]{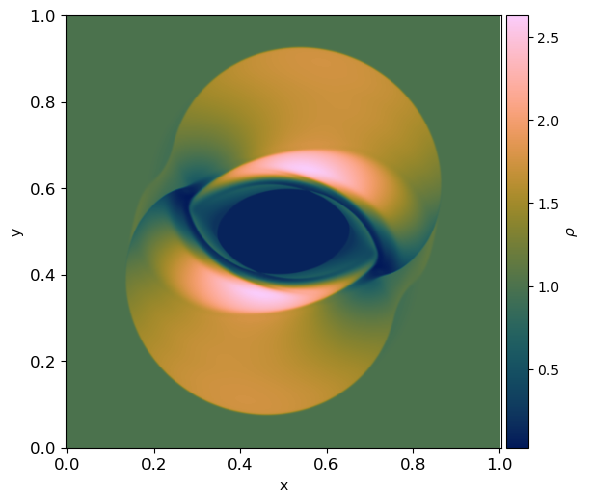}
  \includegraphics[width=0.5\columnwidth]{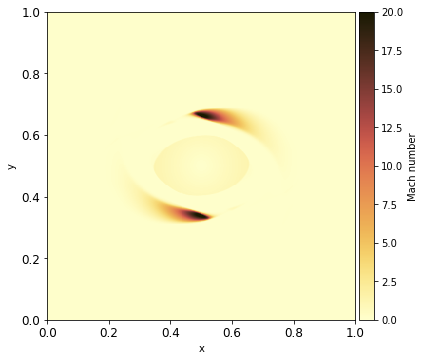}
  \includegraphics[width=0.5\columnwidth]{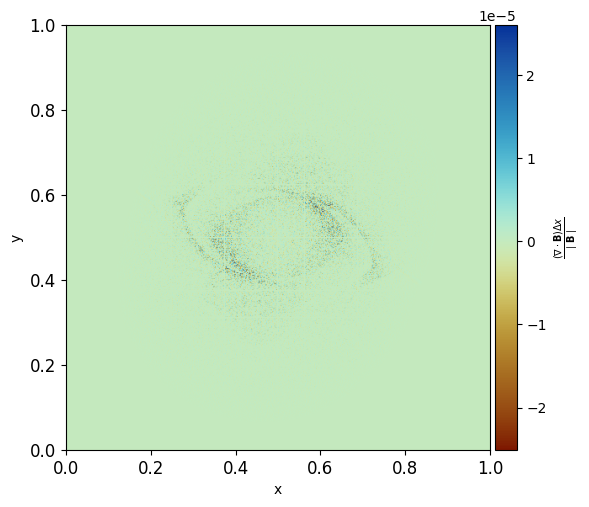}
  \caption{Magnetic rotor test. From left to right: density, magnetic pressure, Mach number and $\frac{\nabla \cdot \bf{B}}{\mid \bf{B} \mid}$. See text for more details.}
  \label{fig:mag_rotor}
\end{figure*}

\subsection{Magnetic rotor}
\label{subsec:magnetic_rotor}
The magnetic rotor tests the propagation of strong torsional Alf\'en waves as well as solver behaviour in moderately high Mach numbers (Ma = 20). A dense disc of fluid rotates within a static fluid background, with a gradual velocity tapering layer between the disc edge and the ambient fluid. An initially uniform magnetic field is present, which is twisted with the disc rotation. The magnetic field is strong enough that it wraps around the rotor diminishing the rotor's angular momentum. The increased magnetic pressure around the rotor compresses the fluid in the rotor, giving it an oblong shape. The experimental setup is introduced by \cite{Balsara_1999}, but we use a more stringent variant of it from \cite{Guillet_2019}, which is summarised below.

The simulation is in two dimensional square, $0 \leq x, y \leq 1$ (512 $\times$ 512 cells) with periodic boundary conditions everywhere. The gas pressure and magnetic fields are uniform in the whole domain, with $P_{gas} = 1$, $\gamma=1.4$ and $\mathbf{B} = (5/\sqrt{4\pi},0,0)$. The gas density is

\begin{equation}
    \rho =
    \left\{
    \begin{array}{l}
    10 \quad \quad \quad \rm{if} \ r < r_0    \\
    1 + 9f \quad \ \rm{if} \ r_0 \leq r \leq r_1  \\
    1 \quad \quad \quad \ \rm{if} \ r_1 < r \\
    \end{array},
    \right.
\end{equation}
where $r_0=0.1$, $r_1 = 0.115$, $r$ is the radial distance from the centre of the box $c$ and $f= (r_1 - r)/(r_1 - r_0)$ is the tapering function for the taper region between the disc and the background. Velocities are
\begin{equation}
    (u_x, u_y) =
    \left\{
    \begin{array}{l}
    (\frac{v_0(c-y)}{r_0}, \frac{v_0(x-c)}{r_0}) \quad \quad \quad \rm{if} \ r < r_0    \\
    (f\frac{v_0(c-y)}{r_0}, f\frac{v_0(x-c)}{r_0}) \quad \ \rm{if} \ r_0 \leq r \leq r_1  \\
    (0,0) \quad \quad \quad \quad \quad \ \rm{if} \ r_1 < r \\
    \end{array},
    \right.
\end{equation}
with $v_0=2$. The experiment was run until time $t=0.15$, by which the torsional Alfv\'en waves have almost reached the boundary. Figure \ref{fig:mag_rotor} shows the density $\rho$, magnetic pressure $P_B$, Mach number and the normalised magnetic field divergence, $\frac{\nabla \cdot \bf{B}}{\mid \bf{B} \mid}$, which, unlike in \cite{Guillet_2019}, we do not rescale to cell size $\Delta x$. We can note the very sharp details with no distortions outside the now almond-shaped disc. In the compressed areas the Mach number becomes very high, but it does not pose any issues. The magnetic field is divergence-free within the noise level.

\section{Discussion and conclusions}
\label{sec:discussion}
In this work we presented a new approximate entropy-based HLLD Riemann solver. It works well in both sub- and super-sonic regimes, preserves positive temperature and gas pressure. The numerical tests are very encouraging and indicate that the HLLS solver can be readily used in a wide range of physical conditions and experimental setups.

\subsection{Considerations about $Q_{S}$ being positive definite}
\label{subsubsec:negative_qs}
Even though we demand $\boldsymbol{Q_{S}}$ to be positive definite, in reality some numerical dispersion and diffusion might artificially increase kinetic or magnetic energy, which then in turn breaks total energy conservation, if a corresponding decrease in thermal energy does not happen. Indeed, this is one of the main reasons many Riemann solvers fail in conditions where kinetic or magnetic energy strongly dominate over thermal energy -- the numerical precision of the latter is reduced if it is several orders of magnitude smaller than the other total energy components and it can lead to negative thermal energy. In the Orszag-Tang test, described above in subsection \ref{subsec:OT_vortex}  series of tests, we did check the difference, whether the requirement of  $\boldsymbol{Q_{S}}$ being positive definite is upheld or not. Until time $\approx 0.3$ the difference is not existent (indicating, that numerical dispersion would not artificially decrease the thermal energy in a total energy solver), but later we do observe considerable differences. This difference sometimes could be decisive whether the experiment crashes or not, when total or thermal energy is used as a primary thermodynamic variable.

\subsection{Very low Mach number regimes}
The numerical tests show that HLLS solver can easily handle Mach numbers to as low as 0.01 and with lower values it becomes rather diffusive. Of course, the perturbations at such Mach numbers are on the order of numerical precision (see Gresho vortex in subsection \ref{subsec:gresho_vortex} for details) and can be absolutely indistinguishable from the background. This is very encouraging, as normally Godunov and Roe type Riemann solvers with single precision struggle with Mach numbers below 0.1.
There are different good attempts to modify solvers to go to very low Mach numbers, like adding correction terms into \texttt{star} states, e.g. \cite{Shima_2011,Dellacherie_2016,Minoshima_2021,Chen_2022}, or by using well-balanced schemes. In the latter a hydrostatic equilibrium is imposed directly in the set of dynamic equations, separating primitive variables into equilibrium (stationary) states and dynamical perturbations, as it is done in e.g.  \cite{Greenberg_1996,Hotta_2022,Cuissa_2022}.  This approach has less success and not as flexible, as the first option, although it is indispensable modelling deep atmospheres, as it helps with hydrostatic equilibrium, numerical precision is better when only perturbations are considered, there is nothing really preventing gas in incompressible state to act as compressible gas and ignore the rotational flows. However, these methods are not mutually exclusive. In fact, combining low-dissipation solvers with well-balancing schemes, as shown by \cite{Edelmann_2021}, can be particularly effective for modelling highly subsonic flows in strongly stratified media, such as deep stellar convection, while keeping grid sizes reasonably small.  Our initial attempts to use a simple ad-hoc modification to $u^\star$ and $P_{tot}^\star$ to imitate the transition to incompressible gas at around Mach = 0.2, as in \cite{Minoshima_2021} did not result in any significant improvement and on the contrary, the test experiments presented highly undesirable results, e.g. the Gresho vortex, presented in subsection \ref{subsec:gresho_vortex} had the background gas rotating in the opposite direction. Clearly we need to do more work to balance out the equations if we want to use this method and later expand the formulation to use some form of a well-balanced scheme.

Lastly, developing a fully three-dimensional wave model for strongly interacting states, akin to the approach in \cite{balsara_2015}, is an avenue we would like to explore in the future. While our current HLLS scheme is inherently three-dimensional and unsplit, it still relies on one-dimensional Riemann problems along coordinate-aligned directions. Extending this to a truly multidimensional Riemann solver, incorporating a strongly interacting state that evolves self-similarly and enforces consistency with the full three-dimensional conservation laws, would enable a more isotropic propagation of flow features and improved treatment of entropy generation from irreversible kinetic and magnetic energy dissipation. This would be particularly beneficial in MHD turbulence and low-Mach number flows, where current schemes may suffer from excessive numerical dissipation or directional bias. However, significant work is required to implement such a model, including designing a suitable three-dimensional wave structure and computing numerical fluxes that maintain consistency across multiple interacting Riemann problems.

We will continue our work on very low Mach number regimes as this is rather crucial to the applications we intend to use the solver for.

\subsection{Applications}
The solver was developed with primary intent to use it in the context of Solar, stellar and planetary atmospheres. From the outermost regions, towards the cores, density, temperature and pressure varies by orders of magnitude, but entropy per unit mass stays almost constant. Additionally, HLLS is the preferred solver in situations where magnetic energy and kinetic energy strongly dominate over thermal energy, e.g. Solar chromosphere and corona.

In strongly stratified media, evolving volumetric entropy rather than volumetric energy as a primary variable ensures that entropy gradients, which drive convection, are captured with higher accuracy. This is particularly beneficial in nearly isentropic convection zones, where traditional total-energy-based schemes struggle due to the small thermal energy contrast relative to the total energy. However, even in subadiabatic regions, where entropy gradients are stable and convection is suppressed, the proposed method remains well-behaved. Since entropy is a logarithmic quantity by nature, its variability in subadiabatic regions does not span orders of magnitude, making it a well-conditioned variable for numerical evolution. This helps maintain accuracy in regions where the entropy contrast is small. For internal gravity waves, entropy-based formulations should naturally capture buoyancy oscillations, as entropy perturbations directly relate to restoring forces in a stratified medium. However, the very low Mach numbers typical of subadiabatic regions introduce additional numerical challenges, as discussed above.

We are already employing this solver to simulate the whole Solar convective region from 0.655 to 0.998 R$_\odot$ (Popovas et al. in prep) and we can see, that the HLLS solver can maintain the hydrostatic equilibrium much better than a HLLD solver without introducing a well-balanced scheme, as in e.g. \cite{Edelmann_2021,Cuissa_2022}.

\subsection{Future work}
We will keep developing the solver so it would be able to cover a larger range of Mach numbers and in general perform better, e.g.:
\begin{itemize}
    \item solver modifications for very low Mach number regimes;
    \item implementing a well-balanced scheme, consistent with the solver;
    \item solver modifications for relativistic regimes;
    \item heavy optimization to improve memory alignment and vectorization that would be also suitable for modern GPUs;
	\item non-ideal MHD effects, introduce entropy generation in Ohmic dissipation and ambipolar diffusion.
\end{itemize}

\begin{acknowledgements}
This research was supported by the Research Council of Norway through its Centres of Excellence scheme, project number 262622, and through grants of computing time from the Programme for Supercomputing, as well as through the Synergy Grant number 810218 (ERC-2018-SyG) of the European Research Council. Lastly, we are grateful to the anonymous referee for in-depth review, detailed and insightful comments, and useful suggestions, that significantly improved the manuscript.
\end{acknowledgements}

\bibliography{ms}

\begin{thebibliography}{53}
\expandafter\ifx\csname natexlab\endcsname\relax\def\natexlab#1{#1}\fi

\bibitem[{Abel(2011)}]{abel_2011}
Abel, T. 2011, Monthly Notices of the Royal Astronomical Society, 413, 271

\bibitem[{Balsara(2015)}]{balsara_2015}
Balsara, D.~S. 2015, Journal of Computational Physics, 295, 1

\bibitem[{Balsara \& Spicer(1999)}]{Balsara_1999}
Balsara, D.~S. \& Spicer, D.~S. 1999, Journal of Computational Physics, 149, 270

\bibitem[{Brio \& Wu(1988)}]{Brio1988400}
Brio, M. \& Wu, C. 1988, Journal of Computational Physics, 75, 400

\bibitem[{Bryan {et~al.}(1995)Bryan, Norman, Stone, Cen, \& Ostriker}]{Bryan_1995}
Bryan, G.~L., Norman, M.~L., Stone, J.~M., Cen, R., \& Ostriker, J.~P. 1995, Computer Physics Communications, 89, 149

\bibitem[{{Canivete Cuissa} \& {Teyssier}(2022)}]{Cuissa_2022}
{Canivete Cuissa}, J.~R. \& {Teyssier}, R. 2022, \aap, 664, A24

\bibitem[{{Clarke}(2010)}]{Clarke_2010}
{Clarke}, D.~A. 2010, \apjs, 187, 119

\bibitem[{Dellacherie {et~al.}(2016)Dellacherie, Jung, Omnes, \& Raviart}]{Dellacherie_2016}
Dellacherie, S., Jung, J., Omnes, P., \& Raviart, P.-A. 2016, Mathematical Models and Methods in Applied Sciences, 26, 2525

\bibitem[{{Derigs} {et~al.}(2016){Derigs}, {Winters}, {Gassner}, \& {Walch}}]{Derigs_2016}
{Derigs}, D., {Winters}, A.~R., {Gassner}, G.~J., \& {Walch}, S. 2016, Journal of Computational Physics, 317, 223

\bibitem[{{Edelmann, P. V. F.} {et~al.}(2021){Edelmann, P. V. F.}, {Horst, L.}, {Berberich, J. P.}, {Andrassy, R.}, {Higl, J.}, {Leidi, G.}, {Klingenberg, C.}, \& {Röpke, F. K.}}]{Edelmann_2021}
{Edelmann, P. V. F.}, {Horst, L.}, {Berberich, J. P.}, {et~al.} 2021, A\&A, 652, A53

\bibitem[{{Evans} \& {Hawley}(1988)}]{Evans_1988}
{Evans}, C.~R. \& {Hawley}, J.~F. 1988, \apj, 332, 659

\bibitem[{{Fromang} {et~al.}(2006){Fromang}, {Hennebelle}, \& {Teyssier}}]{Fromang_2006}
{Fromang}, S., {Hennebelle}, P., \& {Teyssier}, R. 2006, \aap, 457, 371

\bibitem[{{Gallice} {et~al.}(2022){Gallice}, {Chan}, {Loub{\`e}re}, \& {Maire}}]{Gallice_2022}
{Gallice}, G., {Chan}, A., {Loub{\`e}re}, R., \& {Maire}, P.-H. 2022, Journal of Computational Physics, 468, 111493

\bibitem[{{Gardiner} \& {Stone}(2005)}]{Gardiner_2005}
{Gardiner}, T.~A. \& {Stone}, J.~M. 2005, Journal of Computational Physics, 205, 509

\bibitem[{Greenberg \& Leroux(1996)}]{Greenberg_1996}
Greenberg, J.~M. \& Leroux, A.~Y. 1996, SIAM Journal on Numerical Analysis, 33, 1

\bibitem[{Grimm-Strele {et~al.}(2014)Grimm-Strele, Kupka, \& Muthsam}]{Grimm_2014}
Grimm-Strele, H., Kupka, F., \& Muthsam, H. 2014, Computer Physics Communications, 185, 764

\bibitem[{Guillet {et~al.}(2019)Guillet, Pakmor, Springel, Chandrashekar, \& Klingenberg}]{Guillet_2019}
Guillet, T., Pakmor, R., Springel, V., Chandrashekar, P., \& Klingenberg, C. 2019, Monthly Notices of the Royal Astronomical Society, 485, 4209

\bibitem[{Gurski(2004)}]{Gurski_2004}
Gurski, K.~F. 2004, SIAM J. Sci. Comput., 25, 2165–2187

\bibitem[{Happenhofer {et~al.}(2013)Happenhofer, Grimm-Strele, Kupka, Löw-Baselli, \& Muthsam}]{Happenhofer_2013}
Happenhofer, N., Grimm-Strele, H., Kupka, F., Löw-Baselli, B., \& Muthsam, H. 2013, Journal of Computational Physics, 236, 96

\bibitem[{Harten {et~al.}(1983)Harten, Lax, \& Leer}]{Harten_1983}
Harten, A., Lax, P.~D., \& Leer, B.~V. 1983, SIAM Review, 25, 35

\bibitem[{Hawley \& Stone(1995)}]{Hawley_1995}
Hawley, J.~F. \& Stone, J.~M. 1995, Computer Physics Communications, 89, 127, numerical Methods in Astrophysical Hydrodynamics

\bibitem[{Herbin {et~al.}(2020)Herbin, Latch{\'e}, \& Zaza}]{herbin_2020}
Herbin, R., Latch{\'e}, J.-C., \& Zaza, C. 2020, {IMA Journal of Numerical Analysis}, 40, 1792

\bibitem[{Hopkins(2015)}]{Hopkins_2015}
Hopkins, P.~F. 2015, Monthly Notices of the Royal Astronomical Society, 450, 53

\bibitem[{{Hotta} {et~al.}(2022){Hotta}, {Kusano}, \& {Shimada}}]{Hotta_2022}
{Hotta}, H., {Kusano}, K., \& {Shimada}, R. 2022, \apj, 933, 199

\bibitem[{{Ismail} \& {Roe}(2009)}]{ismail_2009}
{Ismail}, F. \& {Roe}, P.~L. 2009, Journal of Computational Physics, 228, 5410

\bibitem[{{Li}(2005)}]{Li_2005}
{Li}, S. 2005, Journal of Computational Physics, 203, 344

\bibitem[{{Londrillo} \& {Del Zanna}(2000)}]{Londrillo_2000}
{Londrillo}, P. \& {Del Zanna}, L. 2000, \apj, 530, 508

\bibitem[{Margolin(2017)}]{Margolin_2017}
Margolin, L. 2017, Entropy, 19

\bibitem[{{McNally} {et~al.}(2012){McNally}, {Lyra}, \& {Passy}}]{McNally_2012}
{McNally}, C.~P., {Lyra}, W., \& {Passy}, J.-C. 2012, \apjs, 201, 18

\bibitem[{Minoshima \& Miyoshi(2021)}]{Minoshima_2021}
Minoshima, T. \& Miyoshi, T. 2021, Journal of Computational Physics, 446, 110639

\bibitem[{{Miyoshi} \& {Kusano}(2005)}]{Miyoshi_2005}
{Miyoshi}, T. \& {Kusano}, K. 2005, Journal of Computational Physics, 208, 315

\bibitem[{Morduchow \& Libby(1949)}]{morduchow_1949}
Morduchow, M. \& Libby, P.~A. 1949, Journal of the Aeronautical Sciences, 16, 674

\bibitem[{{Nordlund} {et~al.}(2018){Nordlund}, {Ramsey}, {Popovas}, \& {K{\"u}ffmeier}}]{Nordlund_2018MNRAS}
{Nordlund}, {\r{A}}., {Ramsey}, J.~P., {Popovas}, A., \& {K{\"u}ffmeier}, M. 2018, \mnras, 477, 624

\bibitem[{{Popovas} \& {J{\o}rgensen}(2016)}]{Popovas_2016}
{Popovas}, A. \& {J{\o}rgensen}, U.~G. 2016, \aap, 595, A130

\bibitem[{{Popovas} {et~al.}(2019){Popovas}, {Nordlund}, \& {Ramsey}}]{Popovas_2019}
{Popovas}, A., {Nordlund}, {\r{A}}., \& {Ramsey}, J.~P. 2019, \mnras, 482, L107

\bibitem[{{Popovas} {et~al.}(2018){Popovas}, {Nordlund}, {Ramsey}, \& {Ormel}}]{Popovas_2018}
{Popovas}, A., {Nordlund}, {\r{A}}., {Ramsey}, J.~P., \& {Ormel}, C.~W. 2018, \mnras, 479, 5136

\bibitem[{{Ramsey} {et~al.}(2012){Ramsey}, {Clarke}, \& {Men'shchikov}}]{Ramsey_2012}
{Ramsey}, J.~P., {Clarke}, D.~A., \& {Men'shchikov}, A.~B. 2012, \apjs, 199, 13

\bibitem[{Roe(1981)}]{Roe1981}
Roe, P. 1981, Journal of Computational Physics, 43, 357

\bibitem[{{Ryu} {et~al.}(1993){Ryu}, {Ostriker}, {Kang}, \& {Cen}}]{Ryu_1993}
{Ryu}, D., {Ostriker}, J.~P., {Kang}, H., \& {Cen}, R. 1993, \apj, 414, 1

\bibitem[{{Salas} \& {Iollo}(1995)}]{Salas_1995}
{Salas}, M.~D. \& {Iollo}, A. 1995, {Entropy jump across an inviscid shock wave}, Final Report Institute for Computer Applications in Science and Engineering, Hampton, VA.

\bibitem[{sheng Chen {et~al.}(2022)sheng Chen, ping Li, Li, Yuan, \& hong Gao}]{Chen_2022}
sheng Chen, S., ping Li, J., Li, Z., Yuan, W., \& hong Gao, Z. 2022, Journal of Computational Physics, 456, 111027

\bibitem[{Shima \& Kitamura(2011)}]{Shima_2011}
Shima, E. \& Kitamura, K. 2011, AIAA Journal, 49, 1693

\bibitem[{Shu \& Osher(1989)}]{Shu1989}
Shu, C.-W. \& Osher, S. 1989, Journal of Computational Physics, 83, 32

\bibitem[{Stone \& Gardiner(2007)}]{Stone_2007}
Stone, J.~M. \& Gardiner, T. 2007, The Astrophysical Journal, 671, 1726

\bibitem[{Stone {et~al.}(2008)Stone, Gardiner, Teuben, Hawley, \& Simon}]{Stone_2008}
Stone, J.~M., Gardiner, T.~A., Teuben, P., Hawley, J.~F., \& Simon, J.~B. 2008, The Astrophysical Journal Supplement Series, 178, 137

\bibitem[{{Teyssier}(2002)}]{Teyssier_2002}
{Teyssier}, R. 2002, \aap, 385, 337

\bibitem[{Teyssier(2015)}]{Teyssier_2015}
Teyssier, R. 2015, Annual Review of Astronomy and Astrophysics, 53, 325

\bibitem[{{Teyssier} {et~al.}(2006){Teyssier}, {Fromang}, \& {Dormy}}]{Teyssier_2006}
{Teyssier}, R., {Fromang}, S., \& {Dormy}, E. 2006, Journal of Computational Physics, 218, 44

\bibitem[{{Thornber} {et~al.}(2008){Thornber}, {Drikakis}, {Williams}, \& {Youngs}}]{Thorbner_2008}
{Thornber}, B., {Drikakis}, D., {Williams}, R.~J.~R., \& {Youngs}, D. 2008, Journal of Computational Physics, 227, 4853

\bibitem[{{Toro}(2019)}]{Toro_2019}
{Toro}, E.~F. 2019, Shock Waves, 29, 1065

\bibitem[{{Toro} {et~al.}(1994){Toro}, {Spruce}, \& {Speares}}]{Toro_1994}
{Toro}, E.~F., {Spruce}, M., \& {Speares}, W. 1994, Shock Waves, 4, 25

\bibitem[{Tóth \& Odstrčil(1996)}]{Toth_1996}
Tóth, G. \& Odstrčil, D. 1996, Journal of Computational Physics, 128, 82

\bibitem[{{Winters} \& {Gassner}(2016)}]{Winters_2016}
{Winters}, A.~R. \& {Gassner}, G.~J. 2016, Journal of Computational Physics, 304, 72

\end{thebibliography}

 \begin{appendix}
\section{The numerical method}
\label{app1:numerical_method}

In Godunov-type schemes the volume-averaged conservative physical quantities at a next time-step are given by integrating an approximate solution to the Riemann problem with left and right states, $U_L$ and $U_R$ at the cell interface. An HLL Rieman  solver \citep{Harten_1983} is constructed by assuming an average intermediate state between the fastest and slowest waves. The information is lost, as slow magneto-acoustic waves are merged together with Alfv\'en and entropy waves. An HLLC \cite{Toro_1994} solver expands and estimates the middle wave of the contact surface. Lastly, HLLD \cite{Miyoshi_2005} expands into four intermediate states. We start with computing the HLL wave speed. Usually the wave speeds are notated by $S_L$ and $S_R$, but to avoid the confusion with entropy, we will note them with $\Sigma_L$ and $\Sigma_R$:
\begin{equation}
\begin{aligned}
    \Sigma_L = \mathrm{min}(u_l,u_r) - \mathrm{max}(c_{f,l}, c_{f,r}) \\
    \Sigma_R = \mathrm{max}(u_l,u_r) + \mathrm{max}(c_{f,l}, c_{f,r})
\end{aligned}
\end{equation}
For convenience, we compute $(\vb{u} \vdot \vb{B})$, which to avoid confusion we denote $\phi$, kinetic energy $E_{kin}$ and magnetic energy $E_{mag}$ for left and right states as well:
\begin{equation}
    \phi_l = u_l A + v_l B_l + w_l C_l ,
\end{equation}
\begin{equation}
    E_{\mathrm{kin},l} = \frac{\rho_l}{2} \left(u_l^2 + v_l^2 + w_l^2 \right) ,
\end{equation}
\begin{equation}
    E_{\mathrm{mag},l} = \frac{1}{2} \left( A^2 + B_l^2 + C_l^2 \right),
\end{equation}
and
\begin{equation}
    \phi_r = u_r A + v_r B_r + w_r C_r ,
\end{equation}
\begin{equation}
    E_{kin,r} = \frac{\rho_r}{2} \left(u_r^2 + v_r^2 + w_r^2 \right) ,
\end{equation}
\begin{equation}
    E_{mag,r} = \frac{1}{2} \left( A^2 + B_r^2 + C_r^2 \right).
\end{equation}
We define the Lagrangian sound speed,
\begin{equation}
    \begin{aligned}
    \upsilon_L = u_l - \Sigma_L \\
    \upsilon_R = \Sigma_R - u_r
    \end{aligned}
\end{equation}
and then compute the acoustic star state:
\begin{equation}
    u^\star = \frac{\rho_r u_r\upsilon_R + \rho_l u_l\upsilon_L + (P_{\mathrm{tot},l}-P_{\mathrm{tot},r})} {\rho_l\upsilon_L + \rho_r\upsilon_R}
\end{equation}
\begin{equation}
    P_{\mathrm{tot}}^\star = \frac{\rho_r\upsilon_R P_{\mathrm{tot},l} + \rho_l\upsilon_l P_{\mathrm{tot},r} + \rho_r\rho_l\upsilon_R\upsilon_L(u_l - u_r)} {\rho_l\upsilon_L + \rho_r\upsilon_R}.
\end{equation}
The left star region variables are
\begin{equation}
    \rho^\star_L = -\frac{\rho_l\upsilon_L}{\Sigma_L-u^\star},
\end{equation}
\begin{equation}
    v^\star_l = \frac{v_l - A B_l(u^\star -u_l)}{e^\star_l},
\end{equation}
\begin{equation}
    w^\star_l = \frac{w_l - A C_l(u^\star -u_l)}{e^\star_l},
\end{equation}
\begin{equation}
    B^\star_l = \frac{B_l e_l}{e^\star_l},
\end{equation}
\begin{equation}
    C^\star_l = \frac{C_l e_l}{e^\star_l},
\end{equation}
where $v_l$ and $w_l$ are left tangential velocities, $A = A_l = A_r$ is the normal component of the magnetic field, $B_l$ and $C_l$ are the left states of the tangential components of the magnetic field,
\begin{equation}
    e^\star_l = -\frac{\rho_l\upsilon_L}{\Sigma_L-u^\star} - A^2,
\end{equation}
and
\begin{equation}
    e_l = -\frac{\rho_l\upsilon_L}{\Sigma_L-u_l} - A^2.
\end{equation}

Entropy is updated as a passive scalar:
\begin{equation}
    S^\star_l = S_l \frac{\rho_l}{\rho^\star_l} \frac{-\upsilon_L}{\Sigma_L-u^\star}.
\end{equation}
Left state Alfv\'en wave speed is
\begin{equation}
    \Sigma_{a,l} = u^\star + \frac{|A|}{\sqrt{\rho^\star_l}}.
\end{equation}
Left star region $(\vb{u} \vdot \vb{B})$, $E_{kin}$ and $E_{mag}$ are
\begin{equation}
    \phi_l^\star = u^\star A + v_l^\star B_l^\star + w_l^\star C_l^\star ,
\end{equation}
\begin{equation}
    E_{\mathrm{kin},l}^\star = \frac{\rho_l^\star}{2} \left(u^{\star 2} + v_l^{\star 2} + w_l^{\star 2} \right) ,
\end{equation}
\begin{equation}
    E_{\mathrm{mag},l}^\star = \frac{1}{2} \left( A^2 + B_l^{\star 2} + C_l^{\star 2} \right).
\end{equation}

Correspondingly, the right star region variables are:
\begin{equation}
    \rho^\star_r = -\frac{\rho_r\upsilon_r}{\Sigma_R-u^\star},
\end{equation}
\begin{equation}
    v^\star_r = \frac{v_r - A B_r(u^\star -u_r)}{e^\star_r},
\end{equation}
\begin{equation}
    w^\star_r = \frac{w_r - A C_r(u^\star -u_r)}{e^\star_r},
\end{equation}
\begin{equation}
    B^\star_r = \frac{B_r e_r}{e^\star_r},
\end{equation}
\begin{equation}
    C^\star_r = \frac{C_r e_r}{e^\star_r},
\end{equation}
where $v_r$ and $w_r$ are right tangential velocities, $A$ is the normal component of the magnetic field, $B_r$ and $C_r$ are the tangential components of the magnetic field,
\begin{equation}
    e^\star_r = -\frac{\rho_r\upsilon_R}{\Sigma_R-u^\star} - A^2,
\end{equation}
and
\begin{equation}
    e_r = -\frac{\rho_r\upsilon_R}{\Sigma_R-u_l} - A^2.
\end{equation}

\begin{equation}
    S^\star_r = S_r \frac{\rho_r}{\rho^\star_r} \frac{-\upsilon_r}{\Sigma_R-u^\star}.
\end{equation}
Right state Alfv\'en wave speed is
\begin{equation}
    \Sigma_{a,r} = u^\star + \frac{|A|}{\sqrt{\rho^\star_r}}.
\end{equation}
Right star region $(\vb{u} \vdot \vb{B})$, $E_{kin}$ and $E_{mag}$ are
\begin{equation}
    \phi_r^\star = u^\star A + v_r^\star B_r^\star + w_r^\star C_r^\star ,
\end{equation}
\begin{equation}
    E_{\mathrm{kin},r}^\star = \frac{\rho_r^\star}{2} \left(u^{\star 2} + v_r^{\star 2} + w_r^{\star 2} \right) ,
\end{equation}
\begin{equation}
    E_{\mathrm{mag},r}^\star = \frac{1}{2} \left( A^2 + B_r^{\star 2} + C_r^{\star 2} \right) .
\end{equation}
Lastly, the double star region variables are:
\begin{equation}
    v^{\star\star} = \frac{v^\star_l\sqrt{\rho^\star_l} + v^\star_r\sqrt{\rho^\star_r} +\Xi(B^\star_r-B^\star_l) }{\sqrt{\rho^\star_l} + \sqrt{\rho^\star_r}},
\end{equation}
\begin{equation}
    w^{\star\star} = \frac{w^\star_l\sqrt{\rho^\star_l} + w^\star_r\sqrt{\rho^\star_r} +\Xi(C^\star_r-C^\star_l) }{\sqrt{\rho^\star_l} + \sqrt{\rho^\star_r}},
\end{equation}
\begin{equation}
    B^{\star\star} = \frac{B^\star_r\sqrt{\rho^\star_l} + B^\star_l\sqrt{\rho^\star_r} +\Xi \sqrt{\rho^\star_l}\sqrt{\rho^\star_r}(v^\star_r-v^\star_l) }{\sqrt{\rho^\star_l} + \sqrt{\rho^\star_r}},
\end{equation}
\begin{equation}
    C^{\star\star} = \frac{C^\star_r\sqrt{\rho^\star_l} + C^\star_l\sqrt{\rho^\star_r} +\Xi \sqrt{\rho^\star_l}\sqrt{\rho^\star_r}(w^\star_r-w^\star_l) }{\sqrt{\rho^\star_l} + \sqrt{\rho^\star_r}},
\end{equation}
where $\Xi$ is a sign function, $\Xi=sign(A)$. Note, that there is no double star variable for entropy. Then double star region $(\vb{u} \vdot \vb{B})$ and $E_{mag}$ are
\begin{equation}
    \phi^{\star\star} = u^\star A + v^{\star\star} B^{\star\star} + w^{\star\star} C^{\star\star} ,
\end{equation}
\begin{equation}
    E_{\mathrm{mag}}^{\star\star} = \frac{1}{2} \left( A^2 + B^{\star\star 2} + C^{\star\star 2} \right) .
\end{equation}
while left and right double star regions for $E_{kin}$ are correspondingly 
\begin{equation}
    E_{\mathrm{kin},l}^{\star\star} = \frac{\rho_l^\star}{2} \left(u^{\star 2} + v^{\star\star 2} + w^{\star\star 2} \right)
\end{equation}
and
\begin{equation}
    E_{\mathrm{kin},r}^{\star\star} = \frac{\rho_r^\star}{2} \left(u^{\star 2} + v^{\star\star 2} + w^{\star\star 2} \right).
\end{equation}
\subsection{Godunov fluxes}
\label{subsec:godunov_flux}
The fluxes are then given by
\begin{equation}
    \boldsymbol{F}_{\mathrm{HLLS}} =
    \left\{
    \begin{array}{l}
    \boldsymbol{F}_L,\ \ \ \ \rm{if} \ \ \Sigma_L > 0 \\
    \boldsymbol{F}^\star_L,\ \ \ \ \rm{if} \ \ \Sigma_L \leq 0 \leq \Sigma_{a,l} \\
    \boldsymbol{F}^{\star\star}_L, \ \ \rm{if} \ \ \Sigma_{a,l} \leq 0 \ \rm{and} \ u^\star > 0 \\
    \boldsymbol{F}^{\star\star}_R, \ \ \rm{if} \ \ u^\star \leq 0 \ \rm{and} \ \Sigma_{a,r} > 0 \\
    \boldsymbol{F}^\star_R,\ \ \ \ \rm{if} \ \ \Sigma_{a,r} \leq 0 \leq \Sigma_R \\
    \boldsymbol{F}_R,\ \ \ \ \rm{if} \ \ \Sigma_R < 0 \\
    \end{array},
    \right.
\end{equation}
where
\begin{equation}
    \boldsymbol{F}_L =
    \left(
    \begin{array}{l}
    \rho_lu_l \\
    \rho_l u^2_l + P_{tot,l} - A^2 \\
    \rho_l u_l v_l - AB_l \\
    \rho_l u_l w_l - AC_l \\
    \rho_l u_l S_l \\
    0 \\
    u_l B_l - A v_l \\
    u_l C_l - A w_l \\
    u_l \left(E_{kin,l} + P_{tot,l}\right) \\
    u_l E_{mag,l} - A\phi_l \\
    u_l \\
    \end{array}
    \right)
\end{equation}
\begin{equation}
    \boldsymbol{F}^\star_L =
    \left(
    \begin{array}{l}
    \rho^\star_l u^\star \\
    \rho^\star_l u^{\star2} + P^\star_{tot} - A^2 \\
    \rho^\star_l u^\star v^\star_l - AB^\star_l \\
    \rho^\star_l u^\star w^\star_l - AC^\star_l \\
    \rho^\star_l u^\star S^\star_l \\
    0 \\
    u^\star B_l^\star - A v_l^\star \\
    u^\star C_l^\star - A w_l^\star \\
    u^\star \left(E_{kin,l}^\star + P_{tot}^\star\right) \\
    u^\star E_{mag,l}^\star - A\phi_l^\star \\
    u^\star \\
    \end{array}
    \right)
\end{equation}
\begin{equation}
    \boldsymbol{F}^{\star\star}_L =
    \left(
    \begin{array}{l}
    \rho^\star_l u^\star \\
    \rho^\star_l u^{\star2} + P^\star_{tot} - A^2 \\
    \rho^\star_l u^\star v^{\star\star} - AB^{\star\star} \\
    \rho^\star_l u^\star w^{\star\star} - AC^{\star\star} \\
    \rho^\star_l u^\star S^\star_l \\
    0 \\
    u^\star B^{\star\star} - A v^{\star\star} \\
    u^\star C^{\star\star} - A w^{\star\star} \\
    u^\star \left(E_{kin,l}^{\star\star} + P_{tot}^\star \right) \\
    u^\star E_{mag}^{\star\star} - A\phi^{\star\star} \\
    u^\star \\
    \end{array}
    \right)
\end{equation}
\begin{equation}
    \boldsymbol{F}^{\star\star}_R =
    \left(
    \begin{array}{l}
    \rho^\star_r u^\star \\
    \rho^\star_r u^{\star2} + P^\star_{tot} - A^2 \\
    \rho^\star_r u^\star v^{\star\star} - AB^{\star\star} \\
    \rho^\star_r u^\star w^{\star\star} - AC^{\star\star} \\
    \rho^\star_r u^\star S^\star_r \\
    0 \\
    u^\star B^{\star\star} - A v^{\star\star} \\
    u^\star C^{\star\star} - A w^{\star\star} \\
    u^\star \left(E_{kin,r}^{\star\star} + P_{tot}^\star \right) \\
    u^\star E_{mag}^{\star\star} - A\phi^{\star\star} \\
    u^\star \\
    \end{array}
    \right)
\end{equation}
\begin{equation}
    \boldsymbol{F}^\star_R =
    \left(
    \begin{array}{l}
    \rho^\star_r u^\star \\
    \rho^\star_r u^{\star2} + P^\star_{tot} - A^2 \\
    \rho^\star_r u^\star v^\star_r - AB^\star_r \\
    \rho^\star_r u^\star w^\star_r - AC^\star_r \\
    \rho^\star_r u^\star S^\star_r \\
    0 \\
    u^\star B_r^\star - A v_r^\star \\
    u^\star C_r^\star - A w_r^\star \\
    u^\star \left(E_{kin,r}^\star + P_{tot}^\star\right) \\
    u^\star E_{mag,r}^\star - A\phi_r^\star \\
    u^\star \\
    \end{array}
    \right)
\end{equation}
\begin{equation}
    \boldsymbol{F}_R =
    \left(
    \begin{array}{l}
    \rho_r u_r \\
    \rho_r u^2_r + P_{tot,r} - A^2 \\
    \rho_r u_r v_r - AB_r \\
    \rho_r u_r w_r - AC_r \\
    \rho_r u_r S_r \\
    0 \\
    u_r B_r - A v_r \\
    u_r C_r - A w_r \\
    u_r \left(E_{kin,r} + P_{tot,r}\right) \\
    u_r E_{mag,r} - A\phi_r \\
    u_r \\
    \end{array}
    \right)
\end{equation}
In the brackets above there are these Godunov fluxes:
\begin{enumerate}
    \item density flux
    \item normal momentum flux
    \item first tangential velocity flux
    \item second tangential velocity flux
    \item entropy flux
    \item normal magnetic field component flux
    \item first tangential magnetic field component flux
    \item second tangential magnetic field component flux
    \item kinetic energy flux
    \item magnetic energy flux
    \item normal velocity
\end{enumerate}

Magnetic field component fluxes are used to update the magnetic fields in order to get the $\frac{dE_{\mathrm{mag}}}{dt}$. If CT method is used to update the magnetic fields, this step is still necessary for consistency (temporarily storing $\frac{dE_{\mathrm{mag}}}{dt}$ separately), as the actual magnetic field update is done through the electromotive force from the cell edges. The kinetic and magnetic energy fluxes are used to compute $\div{\vb{F}_{\mathrm{kin}}}$ and $\div{\vb{F}_{\mathrm{mag}}}$ respectively, while normal velocity (last item) is used to compute $P \left(\div{\vb{u}}\right)$ to get the gas pressure work $W_{gas}$.

\end{appendix}
\end{document}